\documentclass[twocolumn,trackchanges]{aastex701}
\usepackage{booktabs}
\usepackage{hyperref}
\usepackage{overpic}
\usepackage{threeparttable}
\usepackage{graphicx}
\usepackage{float}
\graphicspath{{./}}

\begin{document}

\title{SN~2025wiu: A Slowly Evolving Type Ib Supernova with an Extraordinary Post-maximum Bump}

\correspondingauthor{Ji-an Jiang}
\email{jian.jiang@ustc.edu.cn}

\author[0000-0002-2242-1514,gname=Zhengyan,sname=Liu]{Zhengyan Liu}
\affiliation{Department of Astronomy, University of Science and Technology of China, Hefei 230026, China}
\affiliation{School of Astronomy and Space Sciences, University of Science and Technology of China, Hefei 230026, China}
\email{}

\author[0000-0002-9092-0593]{Ji-an Jiang}
\affiliation{Department of Astronomy, University of Science and Technology of China, Hefei 230026, China}
\affiliation{School of Astronomy and Space Sciences, University of Science and Technology of China, Hefei 230026, China}
\affiliation{National Astronomical Observatory of Japan, 2-21-1 Osawa, Mitaka, Tokyo 181-8588, Japan}
\email{jian.jiang@ustc.edu.cn}

\author[0009-0009-5238-2059]{Zelin Xu}
\affiliation{Department of Astronomy, University of Science and Technology of China, Hefei 230026, China}
\affiliation{School of Astronomy and Space Sciences, University of Science and Technology of China, Hefei 230026, China}
\email{sa21022025@mail.ustc.edu.cn}

\author[0000-0002-1161-9592]{Qiliang Fang}
\affiliation{National Astronomical Observatory of Japan, 2-21-1 Osawa, Mitaka, Tokyo 181-8588, Japan}
\email{qiliang.fang@nao.ac.jp}

\author[0000-0003-2611-7269]{Keiichi Maeda}
\affiliation{Department of Astronomy, Kyoto University, Kitashirakawa-Oiwake-cho, Sakyo-ku, Kyoto 606-8502, Japan}
\email{keiichi.maeda@kusastro.kyoto-u.ac.jp}

\author[0000-0002-1132-1366]{Hanindyo Kuncarayakti}
\affiliation{Tuorla Observatory, Department of Physics and Astronomy, FI-20014 University of Turku, Finland}
\affiliation{Finnish Centre for Astronomy with ESO (FINCA), FI-20014 University of Turku, Finland}
\email{kuncarayakti@gmail.com}

\author[0000-0002-1296-6887]{Lluís Galbany}
\affiliation{Institute of Space Sciences (ICE-CSIC), Campus UAB, Carrer de Can Magrans, s/n, E-08193 Barcelona, Spain}
\affiliation{Institut d'Estudis Espacials de Catalunya (IEEC), 08860 Castelldefels (Barcelona), Spain}
\email{lluisgalbany@gmail.com}

\author[0000-0001-8738-6011]{Saurabh W. Jha}
\affiliation{Department of Physics and Astronomy, Rutgers, The State University of New Jersey, 136 Frelinghuysen Road, Piscataway, New Jersey 08854, USA}
\email{saurabh@physics.rutgers.edu}

\author{\v{Z}eljko Ivezi\'{c}}
\affiliation{Department of Astronomy, University of Washington, Box 351580, Seattle, Washington 98195-1580, USA}
\email{ivezic@uw.edu}

\author[0000-0003-2874-6464]{Peter Yoachim}
\affiliation{Department of Astronomy, University of Washington, Box 351580, Seattle, Washington 98195-1580, USA}
\email{yoachim@uw.edu}

\author[0000-0001-5576-8189]{Andrew J. Connolly}
\affiliation{Department of Astronomy, University of Washington, Box 351580, Seattle, Washington 98195-1580, USA}
\email{ajc@astro.washington.edu}

\author{Dezheng Meng}
\affiliation{Department of Astronomy, University of Science and Technology of China, Hefei 230026, China}
\affiliation{School of Astronomy and Space Sciences, University of Science and Technology of China, Hefei 230026, China}
\email{dezhengmeng@mail.ustc.edu.cn}

\author[0009-0003-9474-8457]{Weiyu Wu}
\affiliation{Department of Astronomy, University of Science and Technology of China, Hefei 230026, China}
\affiliation{School of Astronomy and Space Sciences, University of Science and Technology of China, Hefei 230026, China}
\email{wwyyy@mail.ustc.edu.cn} 

\author[0009-0004-0078-8982]{Ziqing Jia}
\affiliation{Department of Astronomy, University of Science and Technology of China, Hefei 230026, China}
\affiliation{School of Astronomy and Space Sciences, University of Science and Technology of China, Hefei 230026, China}
\email{zqjia@mail.ustc.edu.cn}

\author[0009-0006-0165-6986]{Niko Pyykkinen}
\affiliation{Department of Physics and Astronomy, University of Turku, FI-20014 Turku, Finland}
\email{njpyyk@utu.fi}

\author[0000-0001-8253-6850]{Masaomi Tanaka}
\affiliation{Astronomical Institute, Tohoku University, Aoba, Sendai 980-8578, Japan}
\email{masaomi.tanaka@astr.tohoku.ac.jp}

\author[0000-0001-8537-3153]{Nozomu Tominaga}
\affiliation{National Astronomical Observatory of Japan, 2-21-1 Osawa, Mitaka, Tokyo 181-8588, Japan}
\affiliation{SOKENDAI (The Graduate University for Advanced Studies), Mitaka, Tokyo, 181-8588, Japan}
\email{nozomu.tominaga@nao.ac.jp}

\author{Junhan Zhao}
\affiliation{Department of Astronomy, University of Science and Technology of China, Hefei 230026, China}
\affiliation{School of Astronomy and Space Sciences, University of Science and Technology of China, Hefei 230026, China}
\email{zjunhan@mail.ustc.edu.cn}

\author[0009-0009-2057-7564]{Ming Chen}
\affiliation{Department of Astronomy, University of Science and Technology of China, Hefei 230026, China}
\affiliation{School of Astronomy and Space Sciences, University of Science and Technology of China, Hefei 230026, China}
\email{mingchen26@mail.ustc.edu.cn}

\author[0009-0002-8721-6201]{Tiancheng Lv}
\affiliation{Department of Astronomy, University of Science and Technology of China, Hefei 230026, China}
\affiliation{School of Astronomy and Space Sciences, University of Science and Technology of China, Hefei 230026, China}
\email{tclv@mail.ustc.edu.cn}

\author[0000-0002-1517-6792]{Tinggui Wang}
\affiliation{Department of Astronomy, University of Science and Technology of China,
Hefei 230026, China}
\affiliation{School of Astronomy and Space Sciences,
University of Science and Technology of China,
Hefei 230026, China}
\email{twang@ustc.edu.cn}

\author[0000-0002-7835-8585]{Zigao Dai}
\affiliation{Department of Astronomy, University of Science and Technology of China, Hefei 230026, China}
\affiliation{School of Astronomy and Space Sciences, University of Science and Technology of China, Hefei 230026, China}
\email{daizg@ustc.edu.cn}

\author[0000-0002-7660-2273]{Xu Kong}
\affiliation{Department of Astronomy, University of Science and Technology of China, Hefei 230026, China}
\affiliation{School of Astronomy and Space Sciences, University of Science and Technology of China, Hefei 230026, China}
\affiliation{Institute of Deep Space Sciences, Deep Space Exploration Laboratory, Hefei 230026, China}
\email{xkong@ustc.edu.cn}

\author{Luhan Li}
\affiliation{Yunnan Observatories, Chinese Academy of Sciences, Kunming 650216, China}
\affiliation{University of Chinese Academy of Sciences, Beijing 100049, China}
\email{liluhan@ynao.ac.cn}

\author[0000-0002-3231-1167]{Bo Wang}
\affiliation{Yunnan Observatories, Chinese Academy of Sciences, Kunming 650216, China}
\email{wangbo@ynao.ac.cn}

\author{Jianguo Wang}
\affiliation{Yunnan Observatories, Chinese Academy of Sciences, Kunming 650216, China}
\email{wangjg@ynao.ac.cn}

\author[0000-0002-1935-8104]{Yongquan Xue}
\affiliation{Department of Astronomy, University of Science and Technology of China, Hefei 230026, China}
\affiliation{School of Astronomy and Space Sciences, University of Science and Technology of China, Hefei 230026, China}
\email{xuey@ustc.edu.cn}

\author[0000-0002-1330-2329]{Wen Zhao}
\affiliation{Department of Astronomy, University of Science and Technology of China, Hefei 230026, China}
\affiliation{School of Astronomy and Space Sciences, University of Science and Technology of China, Hefei 230026, China}
\affiliation{College of Physics, Guizhou University, Guiyang 550025, China}
\email{wzhao7@ustc.edu.cn}

\author[0000-0003-0694-8946]{Qingfeng Zhu}
\affiliation{Department of Astronomy, University of Science and Technology of China, Hefei 230026, China}
\affiliation{School of Astronomy and Space Sciences, University of Science and Technology of China, Hefei 230026, China}
\email{zhuqf@ustc.edu.cn}

\author[0000-0003-4200-4432]{Lulu Fan}
\affiliation{Department of Astronomy, University of Science and Technology of China, Hefei 230026, China}
\affiliation{School of Astronomy and Space Sciences, University of Science and Technology of China, Hefei 230026, China}
\affiliation{Institute of Deep Space Sciences, Deep Space Exploration Laboratory, Hefei 230026, China}
\affiliation{College of Physics, Guizhou University, Guiyang 550025, China}
\email{llfan@ustc.edu.cn}

\author{Feng Li}
\affiliation{State Key Laboratory of Particle Detection and Electronics, University of Science and Technology of China, Hefei 230026, China}
\email{phonelee@ustc.edu.cn}

\author{Ming Liang}
\affiliation{National Optical Astronomy Observatory (NSF’s National Optical-Infrared Astronomy Research Laboratory) 950 N Cherry Ave. Tucson Arizona 85726, USA}
\email{liangming@gmail.com}

\author{Binyang Liu}
\affiliation{Purple Mountain Observatory, Chinese Academy of Sciences, Nanjing 210023, China}
\affiliation{School of Astronomy and Space Sciences, University of Science and Technology of China, Hefei 230026, China}
\email{byliu@pmo.ac.cn}

\author[0000-0002-3105-3821]{Zhen Wan}
\affiliation{Department of Astronomy, University of Science and Technology of China, Hefei 230026, China}
\affiliation{School of Astronomy and Space Sciences, University of Science and Technology of China, Hefei 230026, China}
\email{zhen_wan@ustc.edu.cn}

\author[0000-0002-4372-0759]{Hairen Wang}
\affiliation{Purple Mountain Observatory, Chinese Academy of Sciences, Nanjing 210023, China}
\affiliation{School of Astronomy and Space Sciences, University of Science and Technology of China, Hefei 230026, China}
\email{hairenwang@pmo.ac.cn}

\author[0000-0003-1617-2002]{Jian Wang}
\affiliation{State Key Laboratory of Particle Detection and Electronics, University of Science and Technology of China, Hefei 230026, China}
\affiliation{Institute of Deep Space Sciences, Deep Space Exploration Laboratory, Hefei 230026, China}
\email{wangjian@ustc.edu.cn}

\author{Ji Yang}
\affiliation{Purple Mountain Observatory, Chinese Academy of Sciences, Nanjing 210023, China}
\affiliation{School of Astronomy and Space Sciences, University of Science and Technology of China, Hefei 230026, China}
\email{jiyang@pmo.ac.cn}

\author[0000-0002-1463-9070]{Hongfei Zhang}
\affiliation{State Key Laboratory of Particle Detection and Electronics, University of Science and Technology of China, Hefei 230026, China}
\email{nghong@ustc.edu.cn}

\author[0000-0003-3728-9912]{Xianzhong Zheng}
\affiliation{Tsung-Dao Lee Institute and Key Laboratory for Particle Physics, Astrophysics and Cosmology, Ministry of Education, Shanghai Jiao Tong University, Shanghai, 201210, China}
\email{xzzheng@sjtu.edu.cn}

% Authors are sorted alphabetically by family name.
% AASTeX v7 author blocks

\begin{abstract}
Stripped-envelope supernovae (SESNe) originate from massive progenitors that have lost most or all of their outer envelopes prior to core collapse. Recent advances in optical time-domain surveys have revealed a growing population of peculiar SESNe with unusual light-curve morphologies, pointing to energy sources beyond the standard radioactive-decay scenario. In this work, we present multi-band photometric and spectroscopic observations of SN~2025wiu, a peculiar Type Ib supernova discovered by the Wide Field Survey Telescope (WFST). SN~2025wiu is characterized by a high luminosity and exceptionally slow photometric evolution, with a prominent post-maximum bump peaking at $\sim50$ days after the primary maximum. Its early light curve also shows a weak excess consistent with shock-cooling emission from a compact progenitor with a radius of $\sim4\,R_\odot$. The spectra show only modest evolution, with persistently weak line features and no narrow emission lines detected even at $\sim110$ days after maximum. By modeling the primary peak and late-time tail with a combination of radioactive decay and additional energy sources, we isolate the post-maximum bump, which has a roughly symmetric profile with a peak luminosity of $\sim2.4 \times 10^{42}~\mathrm{erg\,s^{-1}}$ and a characteristic duration of $\sim42$ days. {We discuss circumstellar medium (CSM) interaction, accretion, magnetar activity, and a double $^{56}$Ni distribution as possible origins of the bump. The latter three scenarios face challenges in reproducing the post-maximum bump, while delayed CSM interaction remains a possible explanation.}
% While alternative scenarios require additional assumptions or have difficulty reproducing these properties, interaction with an asymmetric circumstellar medium offers a physically viable explanation for the prominent post-maximum bump.

\end{abstract}



\section{Introduction}
Core-collapse supernovae (CCSNe) represent the final evolutionary stages of massive stars with initial masses $\gtrsim 8\,M_\odot$. Prior to core collapse, these massive progenitors can shed their outer envelopes through intense stellar winds \citep{Conti_1976,Smith_2014b}, binary interactions \citep{Podsiadlowski_1992,Sana_2012}, or a combination of both \citep{Fang_2019,Sun_2023}. The varying degree of envelope stripping is directly manifested in the distinct absorption and emission features observed in CCSN spectra. Based on these spectroscopic signatures, stripped-envelope supernovae (SESNe) are classified into Types~IIb, Ib, and Ic, which correspond to the progressive loss of the hydrogen envelope, the entire hydrogen envelope, and both the hydrogen and helium envelopes, respectively.

Historically, the light curves of standard SESNe are well-understood to be primarily powered by the radioactive decay chain of ${}^{56}\mathrm{Ni} \rightarrow {}^{56}\mathrm{Co} \rightarrow {}^{56}\mathrm{Fe}$ \citep{Arnett_1980,Arnett_1982}. On much shorter timescales, the cooling of shock-heated ejecta (i.e., shock breakout and subsequent cooling emission) can also shape the early-time light curves within the first few days post-explosion, offering crucial constraints on the progenitor's radius and envelope structure \citep{Matzner_1999,Soderberg_2008,Rabinak_2011}. {However, an increasing number of modern, high-cadence transient surveys have revealed a growing sample of SESNe whose overall light-curve evolution, especially around or after the maximum, cannot be accommodated by these two conventional mechanisms alone \citep{Ertl_2020,Afsariardchi_2021,Woosley_2021}.} For instance, a systematic energy budget analysis of 54 SESNe by \cite{Rodriguez_2024} demonstrated that radioactive decay alone is insufficient to power their full light-curve evolution, establishing a statistical requirement for additional energy sources across the SESN population.

A growing sample of SESNe exhibiting unusual light-curve behaviors has provided important clues regarding these alternative powering mechanisms. Among them, the discoveries of Type~Ibn \citep{Pastorello_2007} and Icn \citep{Gal-Yam_2022} supernovae have established the pivotal role of violent, early-time circumstellar medium (CSM) interaction in shaping the light-curve peak. In addition, late-time or delayed CSM interactions, which occur when the expanding ejecta eventually catch up with more distant material, have been shown to significantly modulate the post-peak evolution of broader SESN subclasses \citep{Sollerman_2020,Kuncarayakti_2023,Pellegrino_2023}. Beyond CSM interaction scenarios, energy injection from a compact remnant left behind by the core collapse represents another widely discussed paradigm. If the remnant is a newborn, highly magnetized neutron star (a magnetar), its magnetic dipole radiation can efficiently deposit energy into the ejecta---a model frequently invoked to explain luminous SESNe \citep{Milisavljevic_2013,Milisavljevic_2018,Wiston_2026} as well as Type~I superluminous supernovae \citep[SLSNe-I;][]{Gal-Yam_2019}. Alternatively, accretion onto the compact object can play a similar power-injecting role; most notably, the recent discovery of periodic modulations in SN~2022jli \citep{Chen_2024a} highlights that episodic mass accretion from a binary companion can act as a viable engine.

In this work, we present observations of SN~2025wiu, a luminous, peculiar Type~Ib supernova. Its light curve exhibits a prominent post-maximum bump that deviates significantly from pure radioactive decay models, strongly suggesting the presence of an additional powering mechanism. SN~2025wiu\footnote{\href{https://www.wis-tns.org/object/2025wiu}{https://www.wis-tns.org/object/2025wiu}} was discovered and publicly reported by the Wide Field Survey Telescope\footnote{\href{https://wfst.ustc.edu.cn}{https://wfst.ustc.edu.cn}} (WFST; \citealt{Wang_2023,TNS_report}) on 2025 August 28 at 21:22 UT (MJD 60915.891) with a magnitude of $g = 22.7 \pm 0.1$\,mag. The latest pre-discovery non-detection was obtained from a WFST $g$-band image on 2025 August 27 at 21:23 UT, which reached a $5\sigma$ limiting magnitude of 23.1\,mag. SN~2025wiu was later classified as a Type~Ib supernova by \cite{TNS_classify} based on a spectrum taken with the Southern African Large Telescope (SALT) equipped with the Robert Stobie Spectrograph (RSS) $\sim 19$\,days after discovery.

\begin{figure}[htbp]
    \centering
    \begin{overpic}[width=0.44\textwidth]{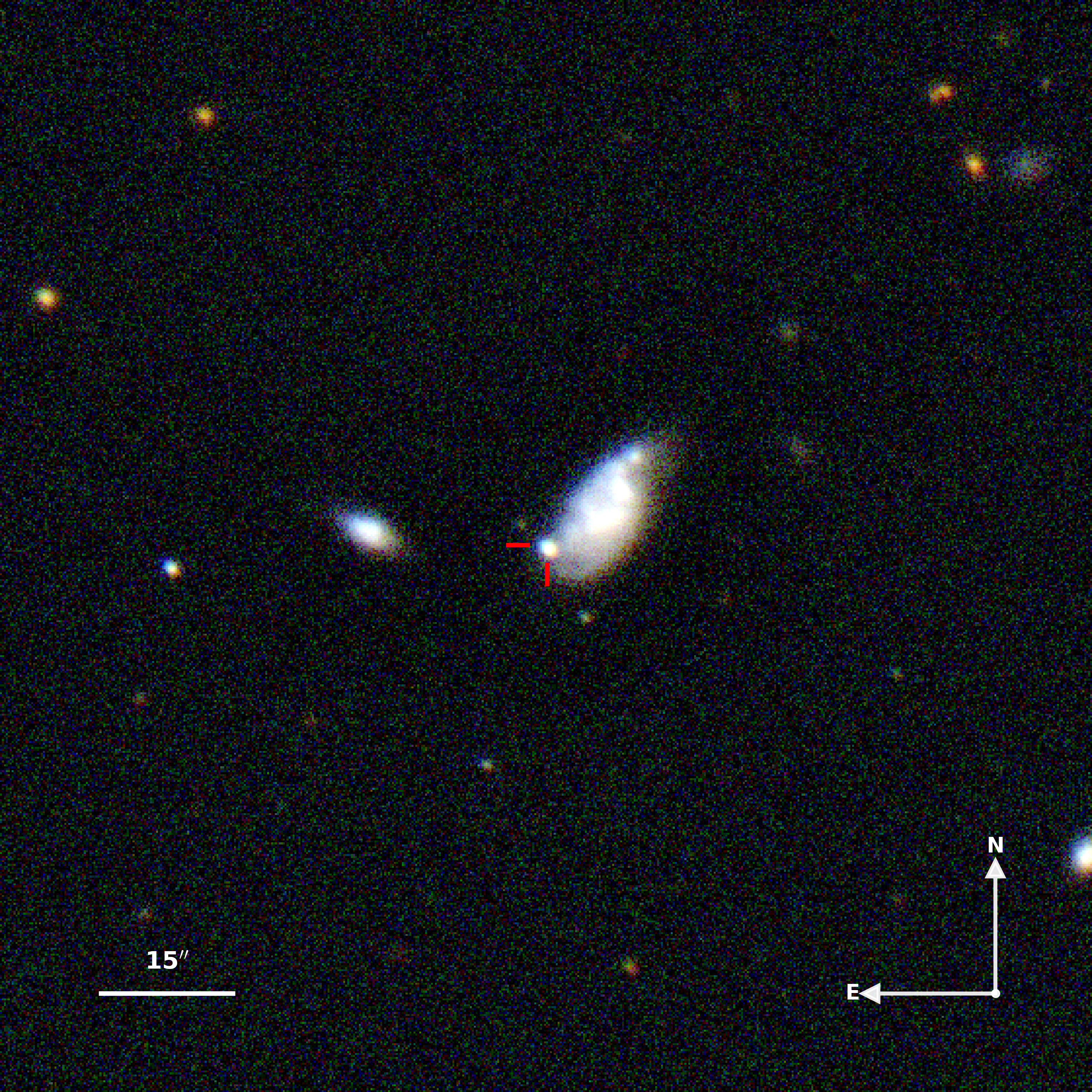}
    \end{overpic}
    \caption{Three-color composite image of SN~2025wiu, observed by the NOT on 2025 November 9. The image is constructed from $g$ (blue), $r$ (green), and $i$ (red) band data. The SN position is marked by a red cross. A 15$^{\prime \prime}$ scale bar is shown in the bottom-left corner.}
    \label{fig:2025wiu_pseudocolor}
\end{figure}

The host galaxy of SN~2025wiu is KUG~0153-006 (shown in Figure~\ref{fig:2025wiu_pseudocolor}), which has a spectroscopic redshift of $0.046927\pm0.000006$ \citep{SDSS_2017}. We adopt a luminosity distance of $215.3$\,Mpc, based on a flat $\Lambda$CDM cosmology with $H_0 = 67.7\,\mathrm{km\,s^{-1}\,Mpc^{-1}}$, $\Omega_{M} = 0.31$, and $\Omega_{\Lambda} = 0.69$ from the Planck 2018 results \citep{Planck2018}. The angular separation between SN~2025wiu and its host galactic center is $\sim 7.1^{\prime \prime}$, corresponding to a projected physical distance of $\sim 6.8$\,kpc. The host galaxy extinction is assumed to be negligible due to the absence of significant \ion{Na}{1}\,D absorption features. The foreground reddening from the Milky Way is determined to be $E(B-V)_{\text{MW}} = 0.028$\,mag \citep{Schlafly_2011}. Therefore, all photometric data were corrected using a total color excess of $E(B-V)_{\text{total}} = 0.028$\,mag, assuming a standard reddening law with $R_V = 3.1$.

This paper is organized as follows. In Section~\ref{sec_2}, we present our observational data and details regarding the data reduction for SN~2025wiu. The photometric and spectroscopic properties are described in Sections~\ref{sec_3} and \ref{sec_4}, respectively. In Section~\ref{sec_5}, we investigate the primary powering mechanisms of SN~2025wiu and model its light-curve evolution. In Section~\ref{sec_6}, we characterize the bump and discuss its possible physical origins. A summary of our main conclusions is given in Section~\ref{sec_7}. Throughout this work, we adopt a reference epoch of MJD~60946.3 as the time zero-point, which corresponds to the rest-frame $r$-band maximum of SN~2025wiu. All phases for comparison SNe are likewise scaled relative to their respective $r$-band maxima. Unless otherwise specified, all upper limits and error bars represent $3\sigma$ and $1\sigma$ confidence levels, respectively.

\section{Observations and data reduction} \label{sec_2}
    \begin{figure*}[htbp]
        \centering
        \begin{overpic}[width=0.95\textwidth]{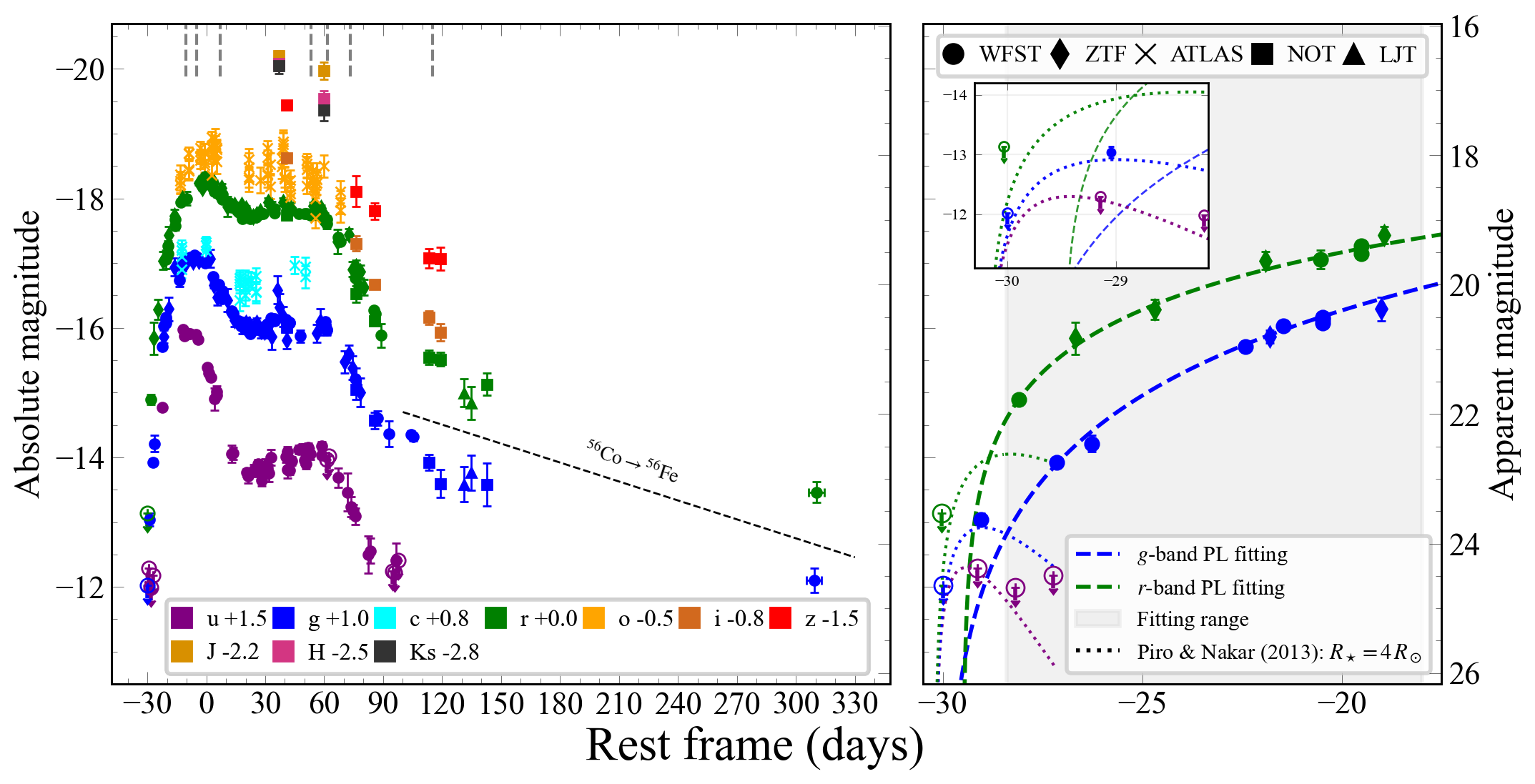}
        \end{overpic}
        \caption{Left: Multi-band light curves of SN~2025wiu (colors represent filters and symbol shapes represent telescopes), corrected for extinction. Open circles denote $3\sigma$ upper limits. The epochs of the spectra presented in this work are also shown as vertical gray dashed lines. The black dashed line indicates the decline rate expected from $^{56}\mathrm{Co}\rightarrow{}^{56}\mathrm{Fe}$ decay under complete trapping ($0.98$ mag per $100$\,days), with an arbitrary vertical offset for comparison. Right: Results of power-law fits to the $g$- and $r$-band early light curves. The time window used for the fit is shown as the gray shaded region. The inset enlarges the early phase from $-30.5$ to $-28$\,days and compares the observations with the shock cooling model (dotted lines) of \cite{Piro_Nakar_2013}. The left and right $y$-axes represent absolute and apparent magnitudes, respectively. The $JHK_s$ magnitudes are converted from the Vega to the AB system using $J_{\rm AB}=J_{\rm Vega}+0.938$, $H_{\rm AB}=H_{\rm Vega}+1.379$, and $K_{s,\rm AB}=K_{s,\rm Vega}+1.900$, following the UKIDSS conversion relations \citep{Hewett_2006}.} 
        \label{fig:multi-band-LC}
    \end{figure*}
    
    \subsection{WFST observations}
    WFST is a 2.5-meter optical telescope with a $6.5\,\mathrm{deg^2}$ field of view and high $u$-band sensitivity \citep{Wang_2023}. It was jointly built by the University of Science and Technology of China (USTC) and the Purple Mountain Observatory (PMO). This specific design enables WFST to conduct deep surveys of the northern sky, making it a powerful instrument for exploring the variable universe and discovering extragalactic transients such as SNe \citep{Hu_2023,Wu_2026}, tidal disruption events, and electromagnetic counterparts of gravitational wave events \citep{Liu_2023}. Operationally, a six-year transient survey project was initiated by WFST in December 2024.

    SN~2025wiu was discovered through the Deep High-cadence $ugr$-band Survey project (``DH$ugr$''; J.~Jiang et al., in preparation), a key project within the WFST formal surveys. The primary observing strategy consisted of daily/hourly-cadence 90\,s exposures in the $u$, $g$, and $r$ bands. Standard data processing was performed using the WFST data pipeline, which is primarily based on the Vera C. Rubin Observatory/LSST image processing pipeline \citep{boschOverviewLSSTImage2018}. The standard LSST pipeline was modified for WFST in two primary ways: first, the addition of extension files to define the CCD configuration, and second, the replacement of the image subtraction module with \texttt{SFFT} to optimize and expedite the transient discovery process \citep{Hu_2022}. We refer the reader to \cite{Cai_2025} for comprehensive details regarding the WFST pipeline.
    
    \subsection{Photometric follow-up and other public data}
    We present 12 epochs of late-time multi-band photometric follow-up observations of SN~2025wiu obtained with the Nordic Optical Telescope (NOT) and the Lijiang 2.4\,m Telescope (LJT; \citealt{2015RAA....15..918F,2019RAA....19..149W}), which are summarized in Table~\ref{tab:pho_summary} in the Appendix. Using the Andalucia Faint Object Spectrograph and Camera (ALFOSC) and the NOT Near-Infrared Camera (NOTCam) on the NOT, we obtained six epochs of $griz$-band and three epochs of $HJK_s$-band photometry spanning $+38$ to $+150$\,days after the observer-frame $r$-band maximum. Additionally, three epochs of $gri$-band photometry were collected with the LJT using the Yunnan Faint-Object Spectrograph and Camera (YFOSC) between $+137$ and $+150$\,days.

    The imaging data from the NOT and LJT were reduced following standard procedures, including bias subtraction and flat-field correction. Astrometric calibration was performed using \texttt{astrometry.net} \citep{astrometry.net}. Individual frames obtained in each band within the same night were stacked using \texttt{SWarp} \citep{SWarp}, and \texttt{Source Extractor} \citep{SExtractor} was employed for source detection. For the optical data, flux calibration was performed using local reference stars from the Pan-STARRS1 (PS1; \citealt{Chambers_2016}) \(3\pi\) survey DR2 catalog \citep{Magnier_2020}. Image subtraction was then carried out using \texttt{HOTPANTS} \citep{hotpants}, with PS1 stacked images \citep{Beck_2021} serving as templates, followed by point-spread-function (PSF) photometry on the subtracted images using \texttt{Photutils} \citep{Photutils}. For the near-infrared \(JHK_s\) data, photometric calibration was instead performed using reference stars from the UKIDSS LAS DR9 catalog \citep{Lawrence_2006}. Although the NOTCam $K_s$ and UKIDSS $K$ filter response curves differ somewhat, this difference does not affect the overall trend of the light-curve evolution. No image subtraction was applied to the near-infrared images. For epochs in which the SN was clearly detected above the underlying host-galaxy background, PSF photometry was performed directly on the sky-background-subtracted science images using \texttt{Photutils}, with the local background estimated around the target. Epochs in which the SN could not be reliably distinguished from the background were excluded from this procedure. 

    We also queried the Zwicky Transient Facility \citep[ZTF;][]{Bellm_2019,graham_zwicky_2019} forced photometry service\footnote{\href{https://ztfweb.ipac.caltech.edu/cgi-bin/requestForcedPhotometry.cgi}{https://ztfweb.ipac.caltech.edu/cgi-bin/requestForcedPhotometry.cgi}} \citep{ztf_force} and the Asteroid Terrestrial-impact Last Alert System \citep[ATLAS;][]{Tonry_2018,Smith_2020} forced photometry server\footnote{\href{https://fallingstar-data.com/forcedphot}{https://fallingstar-data.com/forcedphot}} \citep{ATLAS_force} to harvest supplementary photometric data. To ensure data quality, a signal-to-noise ratio (SNR) $ > 4$ and a magnitude uncertainty $< 0.3$\,mag were required to select valid detections.

    The comprehensive multi-band ($ugcroizJHK_{\rm s}$) light curves are shown in Figure~\ref{fig:multi-band-LC}, providing complete evolutionary coverage and revealing a prominent post-maximum bump. The $g$- and $r$-band photometry at $\sim300$\,days after maximum was obtained by stacking multiple exposures before image subtraction. The absolute magnitudes have been corrected for the extinction. No K-correction is applied. All compiled photometric data are listed in Table~\ref{tab:pho_data_list} of the Appendix.

\subsection{Spectroscopy}
    \begin{figure*}[htbp]
        \centering
        \begin{overpic}[width=0.95\textwidth]{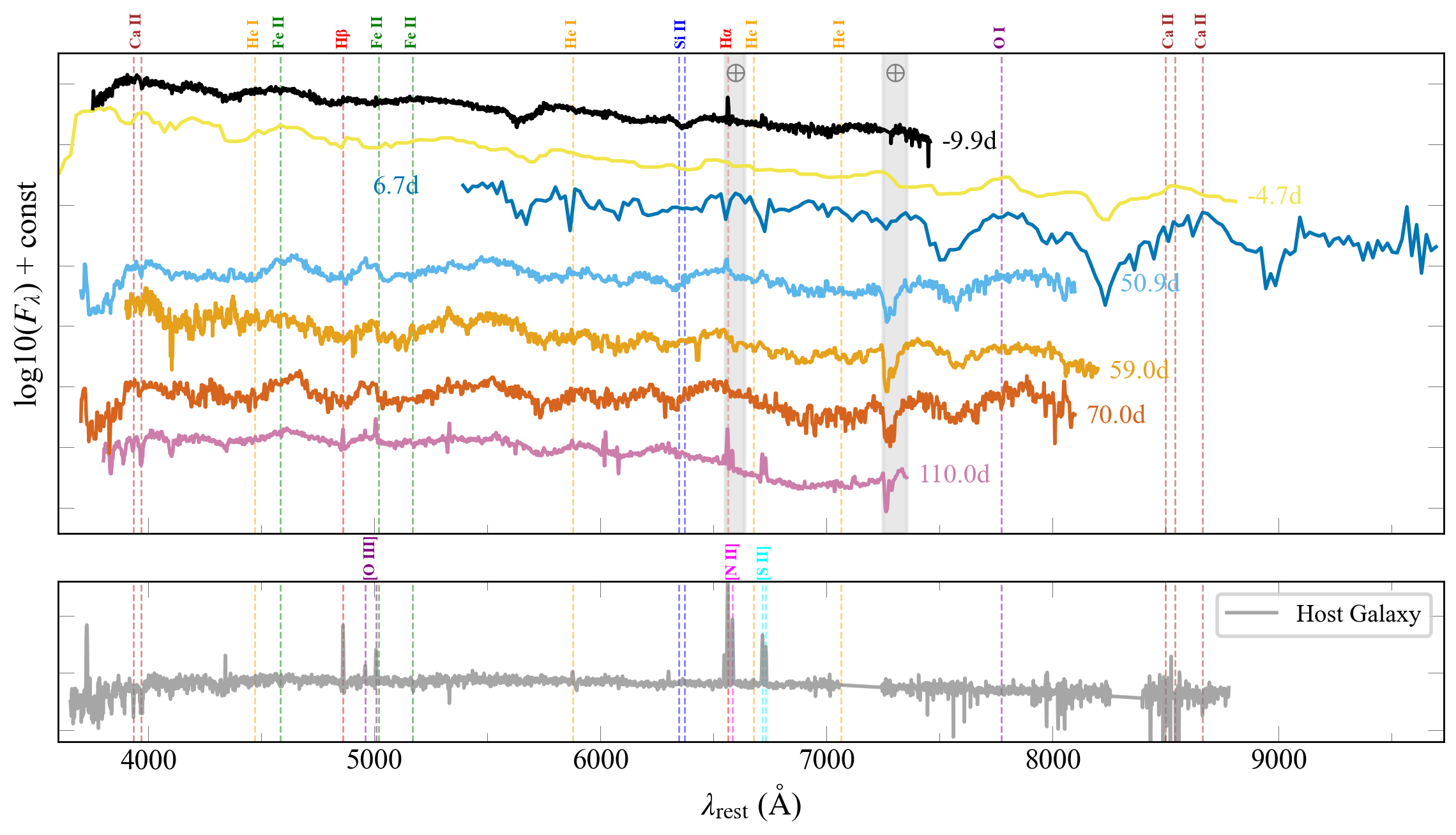}
        \end{overpic}
        \caption{Top: Spectra of SN~2025wiu obtained at multiple epochs with different instruments. For display purposes, an arbitrary offset has been applied to the flux of each spectrum. Telluric features are marked with gray lines. Bottom: The spectrum of the host galaxy. The locations of some typical spectral lines at zero velocity are indicated. The $+6.7$-day P200 spectrum has a limited SNR and is therefore rebinned for display; the narrow absorption features near 5900 and 6700\,\AA\ may be noise artifacts.}
        \label{fig:2025wiu_spec_seq}
    \end{figure*}

    Seven spectral follow-up observations were carried out for SN~2025wiu, spanning $-10$ to $+110$\,days, as summarized in Table~\ref{tab:spec_summary} of the Appendix. Three spectra were collected around the light-curve maximum: these include one publicly available spectrum retrieved from the Transient Name Server (TNS; \citealt{P60_TNS}) and two others obtained with SALT/RSS \citep{2006SPIE.6269E..2AS} and the Next Generation Palomar Spectrograph (NGPS) on the Palomar 200-inch Telescope (P200). Around the post-maximum bump phase, three spectra were taken using LJT/YFOSC and the Kitt Peak Ohio State Multi-Object Spectrograph (KOSMOS) on the Astrophysical Research Consortium (ARC) 3.5\,m telescope. Additionally, one late-time spectrum after the bump was obtained with the Optical System for Imaging and low-Intermediate Resolution Integrated Spectroscopy (OSIRIS) on the Gran Telescopio Canarias (GTC).

    All spectroscopic data were reduced following standard procedures, which included bias subtraction, flat-field correction, spectral extraction, wavelength calibration, and flux calibration using sensitivity functions derived from observations of spectrophotometric standard stars. The GTC/OSIRIS, P200/NGPS, and LJT/YFOSC spectra were reduced using the \texttt{PypeIt} pipeline \citep{pypeit:joss_pub,pypeit:zenodo} with their corresponding instrument configurations. The ARC/KOSMOS spectrum was processed using \texttt{pyKOSMOS} \citep{PyKOSMOS}. The SALT/RSS spectrum was reduced using a custom pipeline based on the PySALT package \citep{2010SPIE.7737E..25C}.

    The full spectroscopic sequence is presented in Figure~\ref{fig:2025wiu_spec_seq}. The spectra obtained around the light-curve maximum show good agreement with those of normal SNe Ib near maximum, based on spectral classification using \texttt{SNID} \citep{SNID}. All spectroscopic data are available for download via the Weizmann Interactive Supernova Data Repository\footnote{\href{https://www.wiserep.org/}{https://www.wiserep.org/}} (WISeREP).
    
    \subsection{Host Galaxy}
    The properties of the host galaxy of SN~2025wiu were obtained from the results of \texttt{Blast} \citep{jonesBlastWebApplication2024}, a web application that ingests live transient alerts, matches them to host galaxies, and performs photometry on corresponding archival host images. Using the host-galaxy photometry as input, host properties were modeled based on the \texttt{Prospector} \citep{Johnson_2021} Bayesian inference framework. 
    
    The archival host-galaxy images span wavelengths from 150\,nm to 20\,$\mu\mathrm{m}$. According to the fitting results of \texttt{Blast}, the host of SN~2025wiu has a stellar mass of $\log(M_* / M_\odot) = 10.02_{-0.12}^{+0.14}$, a stellar metallicity of $\log(Z_* / Z_\odot) = -0.63_{-0.45}^{+0.22}$, and a star formation rate of $\log(SFR/(M_\odot{\rm yr}^{-1}))=0.84_{-0.25}^{+0.23}$. Compared with the properties of host galaxies of SNe classified as Type Ibc reported in TNS, also extracted from \texttt{Blast}, the overall properties align well with the general population distribution, although the host exhibits a relatively high star formation rate.

\section{Photometric Properties} \label{sec_3}
\subsection{Light-curve Evolution}
    \begin{figure*}[htbp]
        \centering
        \begin{overpic}[width=1.01\textwidth]{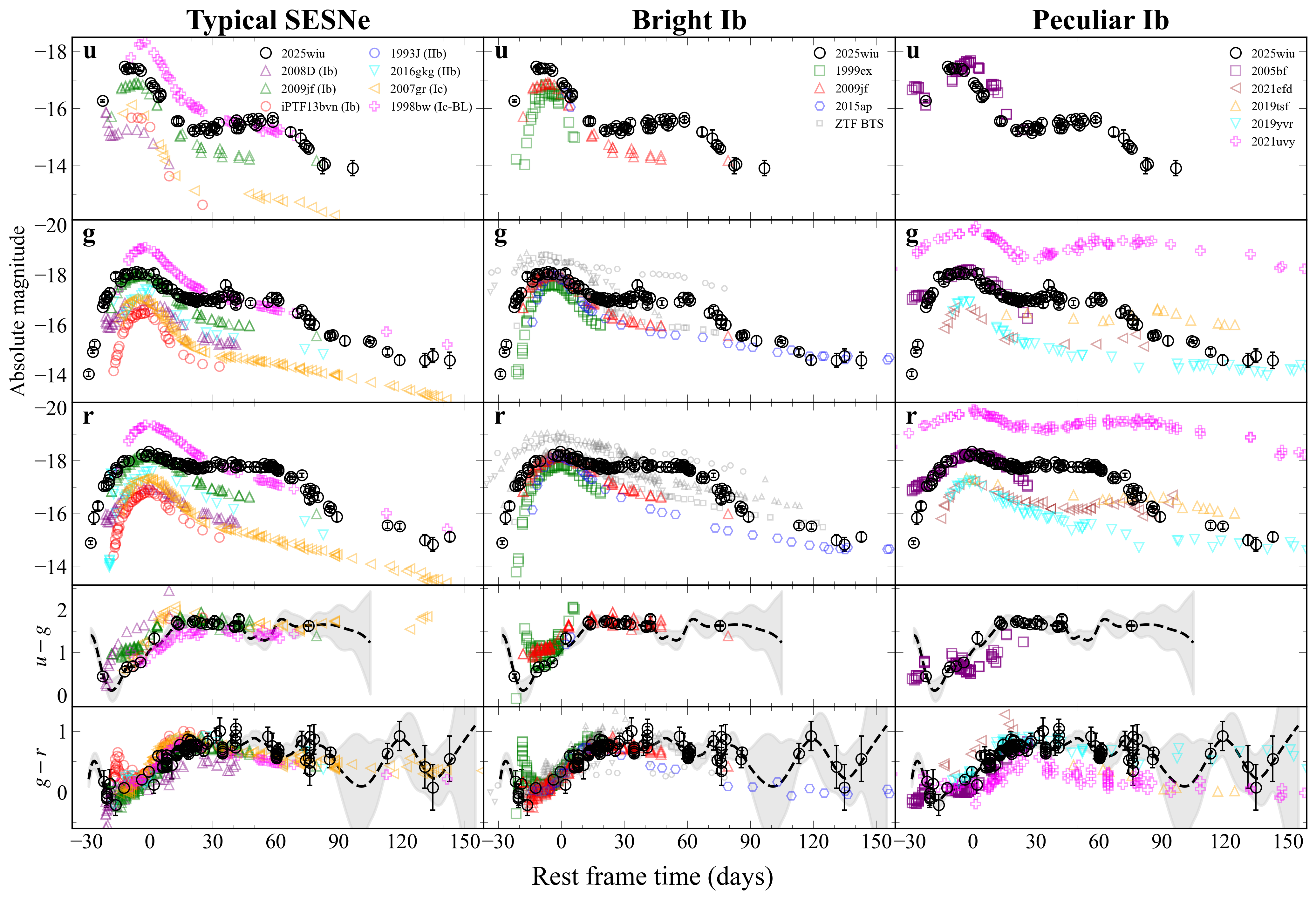}
        \end{overpic}%   <--- 这里的 % 会吃掉换行符
        \caption{Multi-band light curves and color evolution of SN~2025wiu compared with those of typical SESNe (left), bright SNe Ib (middle), and peculiar SNe Ib (right). All light curves are aligned with the epoch of $r$-band maximum. Black dashed lines show the color evolution of SN~2025wiu derived from Gaussian process (GP) regression. Gray markers in the middle column represent bright SNe Ib with well-sampled light curves from the ZTF BTS sample. For clarity, error bars are shown only for SN~2025wiu.}
        \label{fig:SESN_LC_compare}
    \end{figure*}

    % \begin{figure*}[htbp]
    %     \centering
    %     \begin{overpic}[width=1.01\textwidth]{SESN_LC_compare.png}
    %     \end{overpic}%   <--- 这里的 % 会吃掉换行符
    %     \begin{overpic}[width=0.5\textwidth]{Special_LC_compare.png}
    %     \end{overpic}
    %     \caption{Multi-band light curves of SN~2025wiu compared with those of other SESNe, including well-studied SESNe and bright/peculiar SESNe in the left and right panels, respectively. All light curves are aligned by the epoch of $r$-band maximum. The color evolution trend for SN~2025wiu is fitted using GP regression, as denoted by the black dashed line. Some bright SNe Ib with complete light-curve evolution from the ZTF BTS sample are also represented by gray markers in the right panel. For clarity of comparison, error bars are displayed only for SN~2025wiu and are not shown for the other SNe.}
    %     \label{fig:SESN_LC_compare}
    % \end{figure*}

    As shown in the left panel of Figure~\ref{fig:multi-band-LC}, rising from an initial detection of $M_g = -14.0\pm0.1$\,mag, SN~2025wiu reached peak absolute magnitudes of $M_u = -17.5\pm0.1$\,mag, $M_g = -18.11\pm0.01$\,mag, and $M_r = -18.24\pm0.04$\,mag. These peak values are derived via Gaussian Process (GP) regression utilizing a Mat\'{e}rn-$3/2$ kernel. The explosion epoch is estimated to be $-29.8\pm0.6$\,days relative to the reference epoch ($r$-band maximum), as constrained by a power-law (PL) fit to the early-phase $g$- and $r$-band data. Compared to the best-fit PL model, a clear flux excess is present in the first $g$-band detection. Given the luminosity and the concurrent $u$-band non-detection, this early excess likely originates from shock cooling emission rather than a shock breakout signal. To examine whether shock cooling from a compact progenitor can account for this feature, we compare the early photometry with the analytic model of \citet{Piro_Nakar_2013}. We adopt a progenitor radius $R_\star=4\,R_\odot$, ejecta mass $M_{\rm ej}=2\,M_\odot$, explosion energy $E=10^{51}\,\mathrm{erg}$, and opacity $\kappa=0.2\,\mathrm{cm^2\,g^{-1}}$, appropriate for fully ionized hydrogen-poor ejecta. The explosion epoch is fixed at $-30.2$\,days relative to the observed maximum. The resulting WFST $ugr$ light curves are calculated by approximating the spectral energy distribution as a blackbody at the model color temperature. This model is used as a representative comparison rather than a formal fit to the early photometry.
    
    To search for any potential high-energy shock breakout signatures, we inspected the archival data from the Wide-field X-ray Telescope onboard the Einstein Probe (EP-WXT; \citealt{Yuan_2018}). Within the estimated window of the explosion epoch ($-29.8\pm 0.6$\,days), eight observations with exposure times ranging from 1 to 3\,ks covered the position of SN~2025wiu. Together, these observations covered approximately 15\% of the estimated explosion-time window. However, no transient X-ray source was detected within the positional error circle of EP-WXT. For comparison, the SN~2008D X-ray flash reached a peak luminosity of $\sim6\times10^{43}$\,erg\,s$^{-1}$ in the 0.3--10\,keV band \citep{Soderberg_2008}. At the distance of SN~2025wiu, this corresponds to a peak flux of $\sim7\times10^{-12}$\,erg\,s$^{-1}$\,cm$^{-2}$ after conversion to the 0.5--4\,keV band of EP-WXT using the spectral model of SN~2008D, below the typical sensitivity ($\sim2\times10^{-11}$\,erg\,s$^{-1}$\,cm$^{-2}$) of EP-WXT for a 1\,ks exposure.
    
    Following the light-curve maximum, SN~2025wiu shows a significant re-brightening phase starting at $\sim 23$\,days, which is particularly prominent in the $u$-band light curve. Over the course of a $\sim 30$-day rise, it climbed to a second peak with absolute magnitudes of $M_u = -15.6\pm0.1$\,mag, $M_g = -17.17\pm0.04$\,mag, and $M_r = -17.88\pm0.05$\,mag, before entering a rapid decline phase. It is worth noting that, during the development of this prominent bump, subtle secondary fluctuations were also captured in the $g$ and $r$ bands. At late phases ($\gtrsim 75$\,days), SN~2025wiu continued to display extended, smaller-scale bumpy features in both the $g$ and $r$ bands. From 100 to 300\,days, the $g$- and $r$-band decline rates are comparable to that expected from fully trapped $^{56}{\rm Co}\rightarrow{}^{56}{\rm Fe}$ decay.

    A comparison between SN~2025wiu and other well-studied SESNe is presented in the left panel of Figure~\ref{fig:SESN_LC_compare}; this comparison sample includes the Type~IIb SN~1993J \citep{Richmond_1994}, the broad-lined Type~Ic (Ic-BL) SN~1998bw \citep{Clocchiatti_2011}, the Type~Ic SN~2007gr \citep{Hunter_2009}, and the Type~Ib SNe~2008D \citep{Soderberg_2008}, 2009jf \citep{Valenti_2011}, iPTF13bvn \citep{Cao_2013,Kuncarayakti_2015}, and the Type~IIb SN~2016gkg \citep{berstenSurgeLightBirth2018}. For this comparison, archival $UBVR$-band data were transformed into the $ugr$ bands utilizing the transformation equations of \cite{UBV2ugr}. Although slight differences in filter throughput between the WFST and Sloan Digital Sky Survey (SDSS) $u$ bands could introduce minor systematic offsets, such effects are negligible when analyzing the overall light-curve evolution. Compared to these classic SESNe, SN~2025wiu stands out as a luminous and exceptionally slow-evolving Type~Ib SN. 

    According to the ZTF Bright Transient Survey (BTS) SESN sample compiled to date \citep{Fremling_2020,Perley_2020}, the median $r$-band peak absolute magnitudes are $M_{\mathrm{IIb}} = -17.37 \pm 0.70$\,mag, $M_{\mathrm{Ib}} = -17.39 \pm 0.83$\,mag, $M_{\mathrm{Ic}} = -17.49 \pm 0.82$\,mag, and $M_{\mathrm{Ic-BL}} = -18.46 \pm 1.14$\,mag. Consequently, SN~2025wiu is more luminous than typical Type~IIb, Ib, and Ic SNe, while remaining fainter than the bulk of Type~Ic-BL SNe. The $r$-band rise and decay times (defined from half-peak to peak and peak to half-peak) for SN~2025wiu are 17.4 and 30.4\,days, respectively. Here, the decay timescale was estimated via linear extrapolation because the subsequent re-brightening phase initiated before the flux could fade to half-maximum. For comparison, the median $r$-band rise and decay timescales for SNe Ib within the ZTF BTS sample are $T_{\mathrm{rise,Ib}} = 12.2 \pm 6.7$\,days and $T_{\mathrm{decay,Ib}} = 13.8 \pm 10.2$\,days. The significantly slower decline of SN~2025wiu strongly indicates an early, profound influence of the upcoming secondary bump on the primary light-curve evolution.

    To further investigate the bumpy features of SN~2025wiu, we present a focused comparison with literature samples of luminous and peculiar SNe Ib that exhibit similar light-curve structures in the middle and right panels of Figure~\ref{fig:SESN_LC_compare}, respectively. This comparison sample includes SNe~1999ex \citep{Stritzinger_2002}, 2005bf \citep{Tominaga_2005,2014AJ....147...99M,2019MNRAS.482.1545S}, 2015ap \citep{Prentice_2019,Gangopadhyay_2020}, 2019tsf \citep{Zenati_2025}, 2019yvr \citep{Ferrari_2023}, 2021efd \citep{Pyykkinen_2026}, and 2021uvy \citep{Sharma_2025}. Additionally, several luminous SNe selected from the ZTF BTS sample are represented by gray markers for reference, namely SNe~2019buy, 2019odp, 2019qvt, 2022ujq, 2023epp, 2024gry, and 2025fmr. We also note that while several SESNe exhibiting peculiar light-curve profiles or dramatic re-brightening events have been intensely studied in recent years (e.g., SNe~2022jli, \citep{Chen_2024a}; 2022xxf, \citep{Kuncarayakti_2023}; 2023aew, \citep{Kangas_2024}), these objects are excluded from this comparison due to distinct spectral classifications or disparate evolutionary timescales. While some morphological similarities are observed between the light curves of SN~2021uvy and SN~2025wiu, the peak luminosity of SN~2021uvy is markedly higher and its overall evolution is slower. With the exception of SN~2021uvy, none of the other known luminous and peculiar SNe Ib display a post-maximum bump comparable in strength to that observed in SN~2025wiu.

    The color evolution of SN~2025wiu and its comparison with other SESNe are also presented in Figure~\ref{fig:SESN_LC_compare}. The $u-g$ and $g-r$ intrinsic colors of SN~2025wiu are derived from same-night photometric observations, with their evolutionary trends captured via GP regression. Prior to the first maximum, following a brief initial phase dominated by shock cooling emission, the color evolution of SN~2025wiu displays an overall red-to-blue trend, closely resembling that of iPTF13bvn. Around the primary peak, the overall color evolution remains entirely consistent with the canonical behavior of other SESNe. During the subsequent re-brightening bump, the colors of SN~2025wiu follow a trajectory similar to the comparison sample but show distinct, subtle fluctuations, particularly in the $g-r$ color index. At late phases ($\gtrsim 80$\,days), these color fluctuations persist and tentatively exhibit a quasi-periodic behavior, albeit with relatively large uncertainties resulting from the sparse sampling cadence.

\subsection{Pseudo-bolometric Light Curve}
    \begin{figure}[htbp]
        \centering
        \begin{overpic}[width=0.47\textwidth]{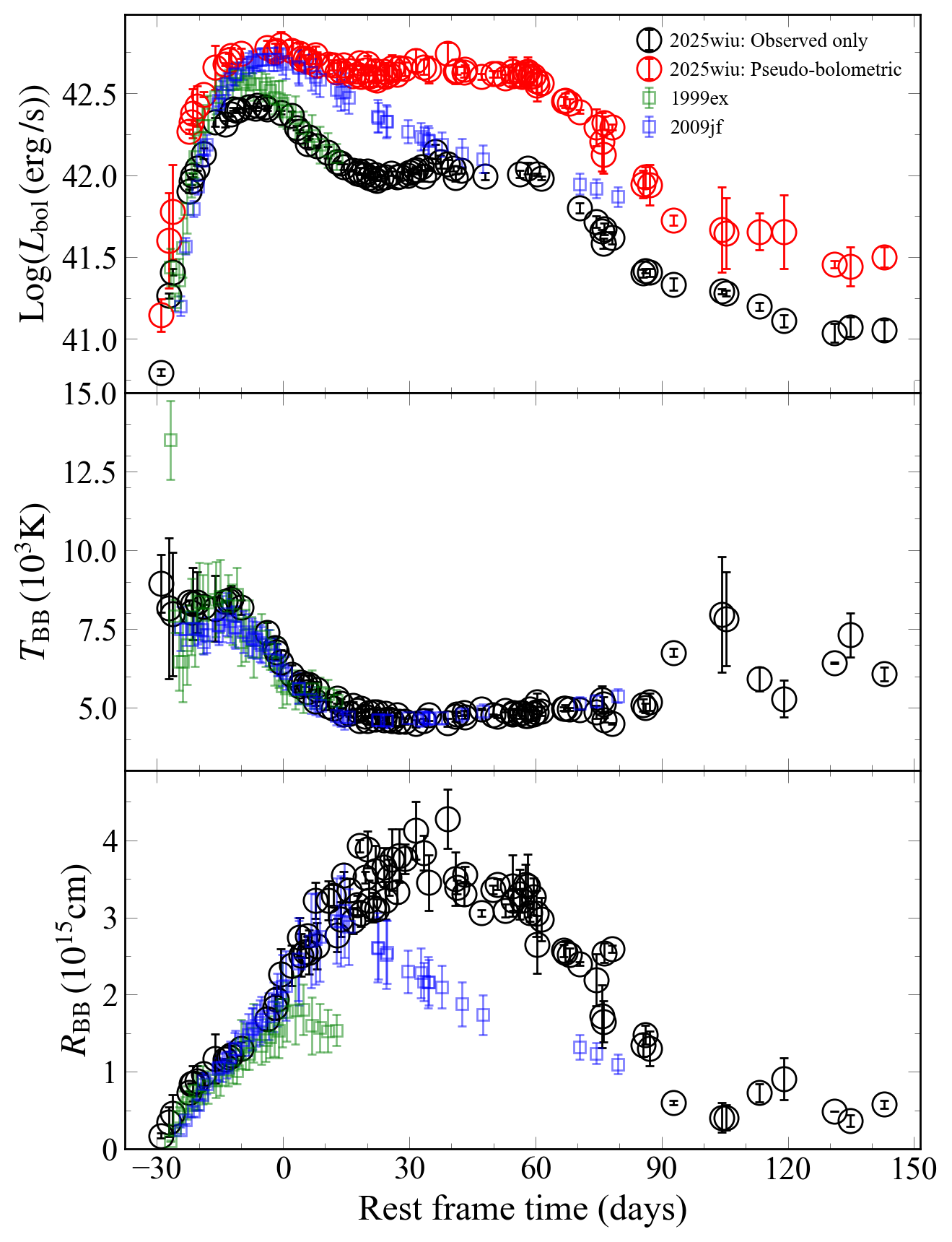}
        \end{overpic}
        \caption{Blackbody fitting results for SN~2025wiu using $ugro$-band data. Top: The evolution of the observed luminosity and the pseudo-bolometric luminosity integrated from 1 to 25000\,\AA. The observed luminosity is derived from the $ugro$ bands due to monitoring completeness. Middle: The evolution of the photospheric temperature. Bottom: The evolution of the photospheric radius. The corresponding evolutions of the pseudo-bolometric luminosity, photospheric temperature, and photospheric radius of SNe 1999ex (green) and 2009jf (blue) are also shown for comparison.}
        \label{fig:bolometric_LCs}
    \end{figure}
    The pseudo-bolometric light curve of SN~2025wiu is constructed by fitting the multi-band photometric data with blackbody (BB) functions. Based on monitoring completeness, the fitting is performed via \texttt{Superbol} \citep{Nicholl2018} using the $ugro$ bands, where fluxes for missing phases are estimated through polynomial interpolation. The resulting evolution of the bolometric luminosity, photospheric temperature, and photospheric radius is shown in Figure~\ref{fig:bolometric_LCs}, where the results obtained for SNe 1999ex and 2009jf using \texttt{Superbol} are also shown for comparison. It should be noted that the inferred parameters at late times ($> +90$\,days) should be considered as reference values only, due to the increasing deviation of the spectral energy distribution from a pure BB spectrum. Overall, the pseudo-bolometric light curve shows similar morphological features to the multi-band light curves. The peak pseudo-bolometric luminosity, integrated from 1 to $25000$\,\text{\AA}, is $\log(L_{\mathrm{bol}} / \mathrm{erg\,s^{-1}}) = 42.77 \pm 0.03$ (corresponding to $(5.89\pm0.42)\times10^{42} \,\mathrm{erg\,s^{-1}}$). For comparison, the median peak pseudo-bolometric luminosities $\log(L_{\mathrm{bol}} / \mathrm{erg\,s^{-1}})$ of SESNe integrated over the $4000\text{--}10000$\,\text{\AA} range from \cite{Prentice_2019} are $42.09\pm0.17$ for SNe IIb, $42.2^{+0.4}_{-0.1}$ for SNe Ib, and $42.3^{+0.3}_{-0.2}$ for normal SNe Ic. The band-limited peak luminosity of SN~2025wiu within this identical wavelength range is $\log(L_{\mathrm{bol}}/\mathrm{erg\,s^{-1}})=42.50\pm0.09$, further confirming that SN~2025wiu is a luminous Type~Ib SN. Based on this pseudo-bolometric light curve, a total radiated energy of $\sim 4.1 \times 10^{49}$\,erg was emitted over the $\sim 180$\,day observational baseline.

    As shown in the middle panel of Figure~\ref{fig:bolometric_LCs}, the photospheric temperature initially declined rapidly, consistent with shock-cooling emission, before rising to a maximum of $\sim8500$\,K. It subsequently decreased until $\sim30$\,days, coinciding with the onset of the re-brightening bump and the contraction of the photospheric radius. The expansion velocity of the photosphere is estimated to be $\sim 8000\,\mathrm{km\,s^{-1}}$, as derived from the linear radius evolution between $-29$ and $+10$\,days. During the development of the secondary bump, the photospheric temperature shows a modest increase back to $\sim 5000$\,K, corresponding to the slight decrease in $u-g$ in Figure \ref{fig:SESN_LC_compare} at the same phase. Despite these apparent temperature variations, the BB fits at late phases ($> 90$\,days) may not reflect the intrinsic physical conditions of the SN due to significant deviations from a pure BB spectral energy distribution; therefore, the physical parameters inferred at these late epochs should be regarded as indicative only.

    \begin{figure*}[htbp]
        \centering
        \begin{overpic}[width=0.48\textwidth]{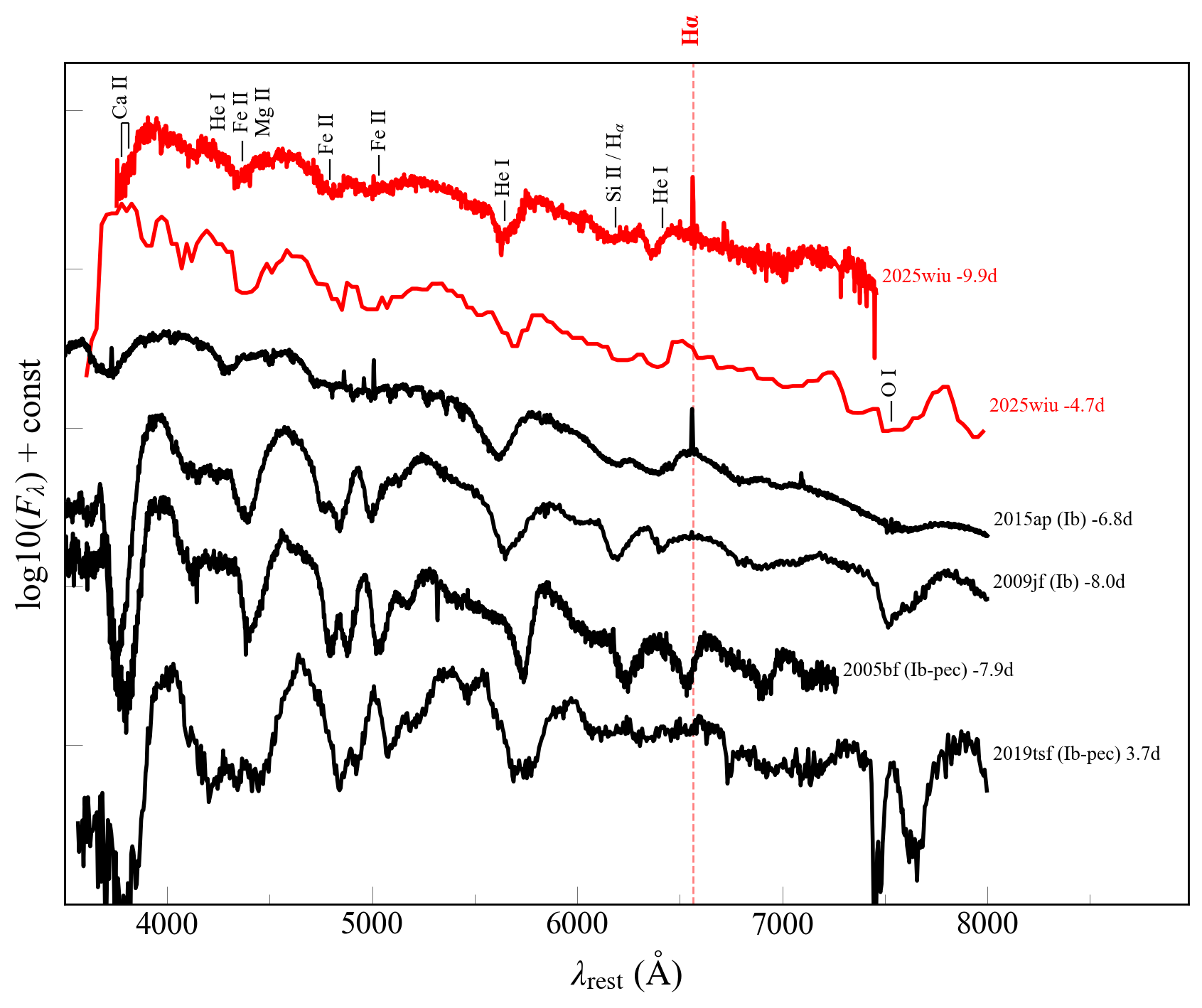}
        \end{overpic}
        \begin{overpic}[width=0.495\textwidth]{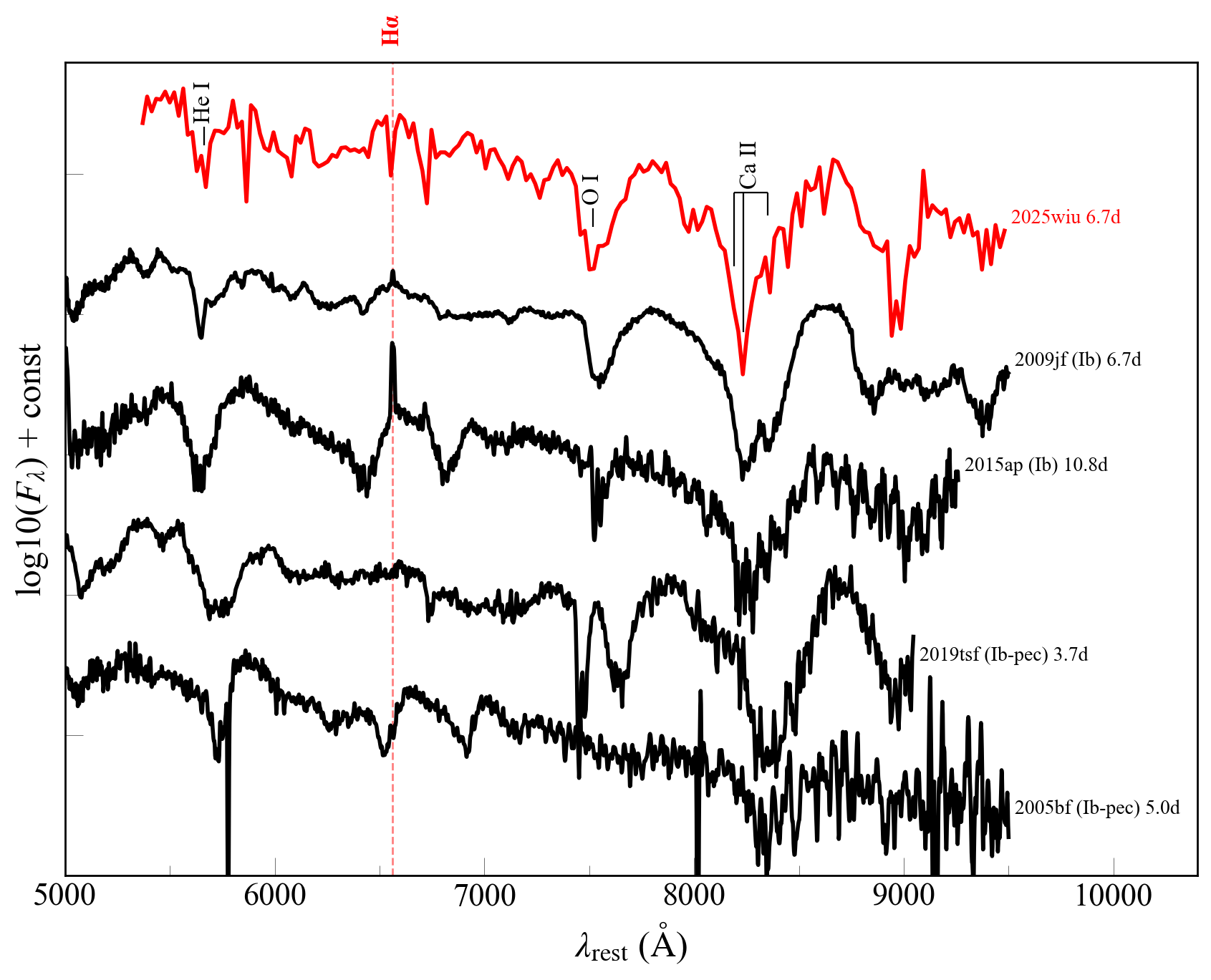}
        \end{overpic}
        \caption{Photospheric-phase spectra of SN~2025wiu (red) and other comparison SESNe (black). Each spectrum is plotted on a logarithmic flux scale and vertically offset for clarity. The main spectral features are labeled, and the blue-shifted positions of selected absorption lines are marked. The $+6.7$ day spectrum has a limited SNR and is therefore rebinned for display; the narrow absorption features near 5900 and 6700\,\AA\ may be noise artifacts.}
        \label{fig:spec_compare_around_max}
    \end{figure*}

    \begin{figure*}[htbp]
        \centering
        \begin{overpic}[width=0.48\textwidth]{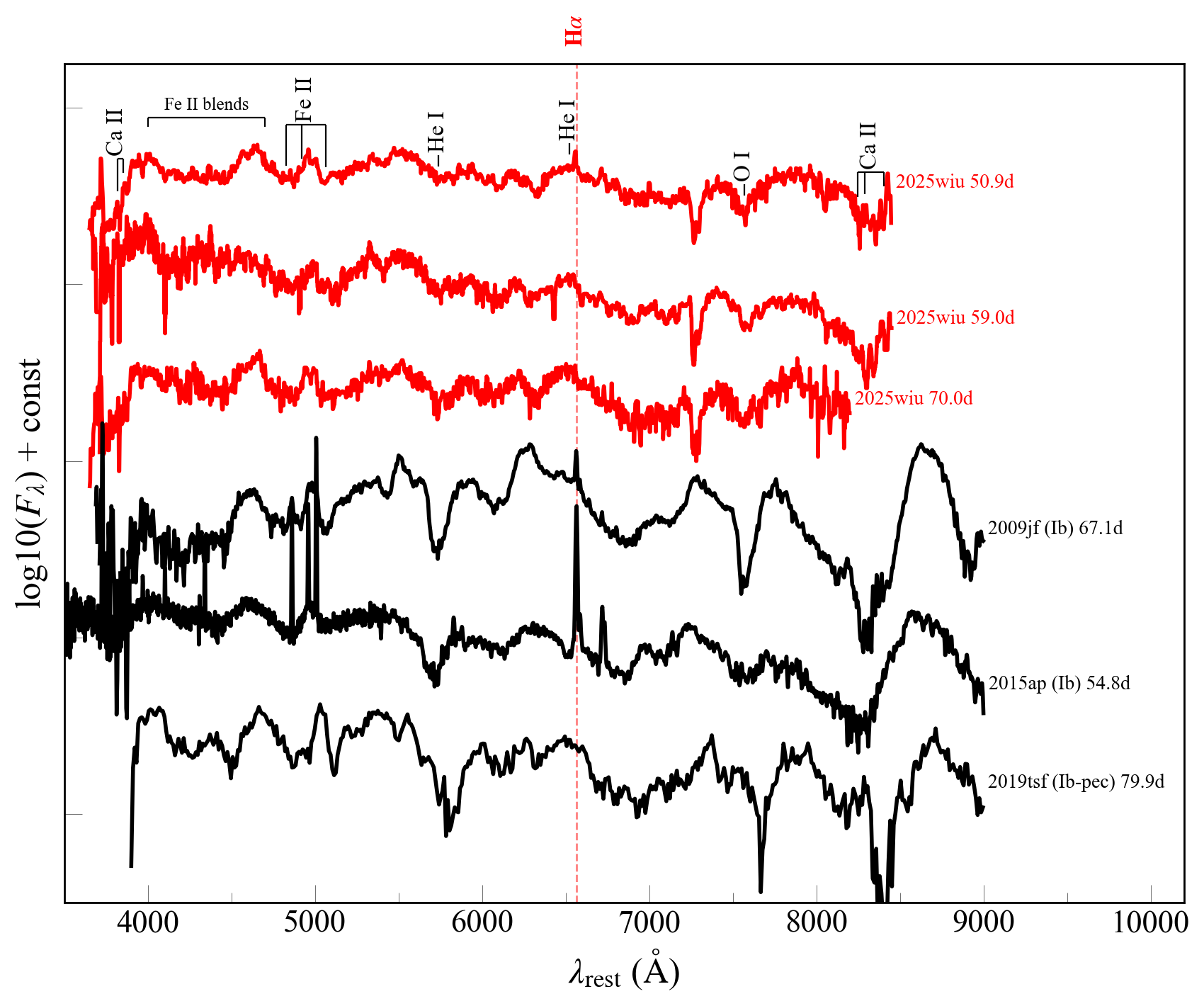}
        \end{overpic}
        \begin{overpic}[width=0.49\textwidth]{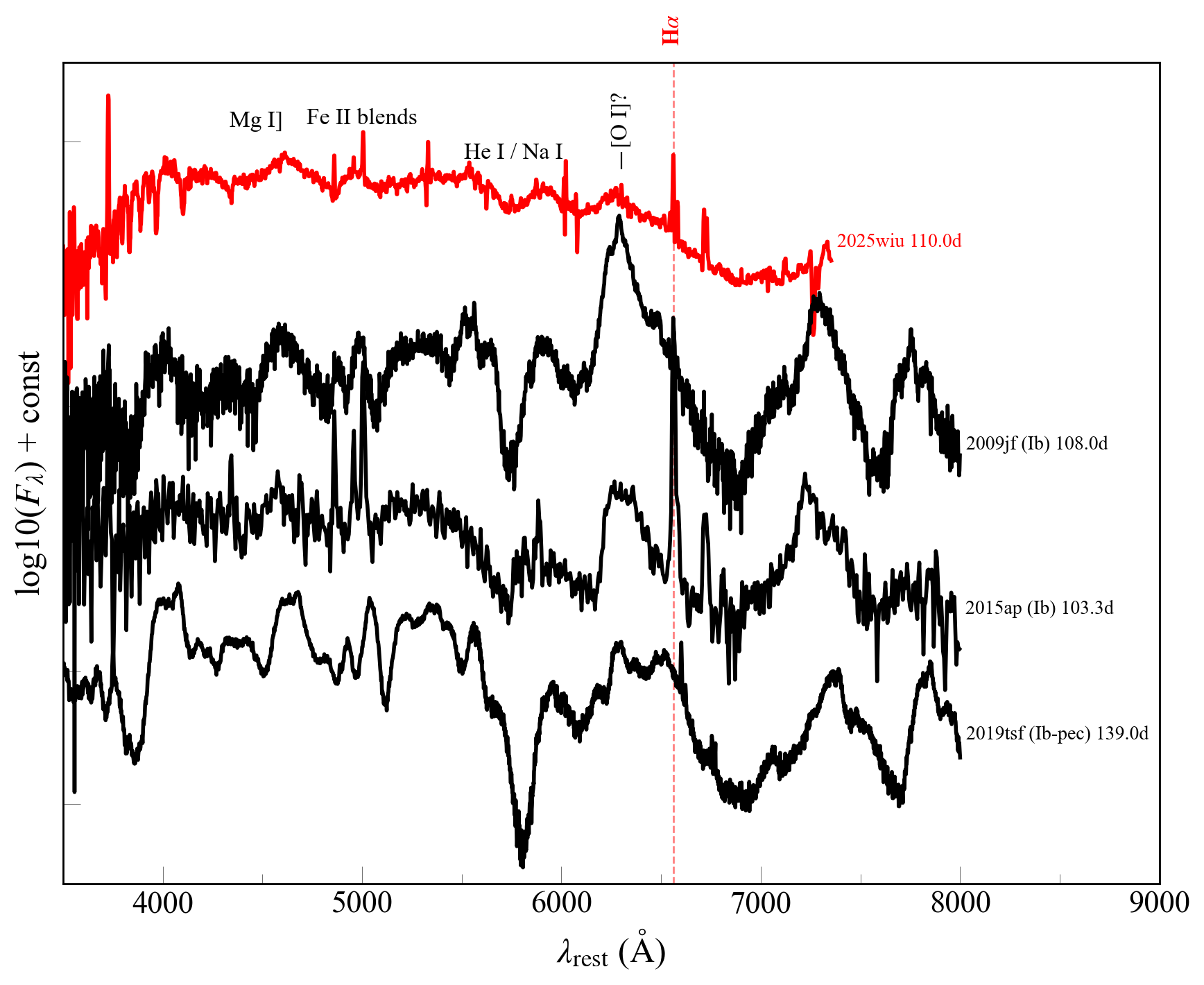}
        \end{overpic}
        \caption{Post-maximum spectra of SN~2025wiu (red) and other comparison SESNe (black). Each spectrum is plotted on a logarithmic flux scale and vertically offset for clarity. The main spectral features are labeled, and the blue-shifted positions of selected absorption lines are marked.}
        \label{fig:spec_compare_post_max}
    \end{figure*}
\section{Spectroscopic properties} \label{sec_4}
    \subsection{Spectral evolution}
    The comprehensive spectral evolution of SN~2025wiu is shown in Figure~\ref{fig:2025wiu_spec_seq}. The narrow Balmer emission lines detected in the spectra at $-9.9$, $+50.9$, and $+110.0$\,days originate from the host galaxy, as confirmed by the accompanying prominent [\ion{O}{3}], [\ion{N}{2}], and [\ion{S}{2}] lines (see also the archival host galaxy spectrum from SDSS \citep{SDSS_2017} in Figure~\ref{fig:2025wiu_spec_seq}). The early-time spectral sequence of SN~2025wiu shows a blue continuum characterized by prominent \ion{He}{1}, \ion{Fe}{2}, and \ion{Ca}{2} features, which is typical of SNe Ib during their early and maximum-light phases. {The expansion velocity estimated from the absorption minimum of the \ion{Fe}{2}~$\lambda5169$ line in the $-9.9$ day spectrum is $\sim8000$ km s$^{-1}$, although the shallow feature introduces substantial uncertainty. The limited wavelength resolution of the $-4.7$ day spectrum prevents a reliable velocity measurement.} Blackbody fits to these $-9.9$ and $-4.7$\,day spectra yielded temperatures of $8100 \pm 200$\,K and $9000 \pm 900$\,K, respectively, which are in agreement with the independent constraints derived from our photometric data.
    
    After the peak, the blue continuum faded and the spectral lines became more prominent and continued to develop. At $+59.0$\,days, the blue continuum showed some enhancement, implying an increase in photospheric temperature, which is also consistent with the color evolution of $u-g$ (see Figure~\ref{fig:SESN_LC_compare}). During the bump, SN~2025wiu showed no significant spectral evolution. In addition, no narrow H or He lines indicative of ejecta--CSM interaction were present during the bump phase, based on spectra without significant host contamination (e.g., the spectra at $+59.0$ and $+70.0$\,days). The weak absorption and emission features in the spectrum at $+110.0$\,days suggest that the SN had not yet fully entered the nebular phase.
        
    \subsection{Spectral comparison among SESNe}
    {The photospheric spectra of SN~2025wiu and our selected SESN comparison sample are shown in Figure~\ref{fig:spec_compare_around_max}, where the blue-shifted positions of the main absorption features are marked. The spectra of SN~2025wiu closely resemble those of the luminous Type~Ib SN~2015ap, particularly in the strengths of the absorption features (e.g., Fe II $\lambda5169$ and He I $\lambda5876$).} The He I \(\lambda5876\) feature is clearly present throughout the photospheric spectral sequence of SN~2025wiu. Interestingly, the \ion{He}{1} profiles in SN~2025wiu are similar to those observed in SNe~2009jf, 2015ap, and 2005bf, but distinctly diverge from the features seen in SN~2019tsf, which also shows post-maximum re-brightening bumps. Considering the disparate luminosities across these objects (see Figure~\ref{fig:SESN_LC_compare}), the specific spectral line profiles at the photospheric phase are likely linked to the peak luminosity.

    \begin{figure}[htbp]
        \centering
        \begin{overpic}[width=0.47\textwidth]{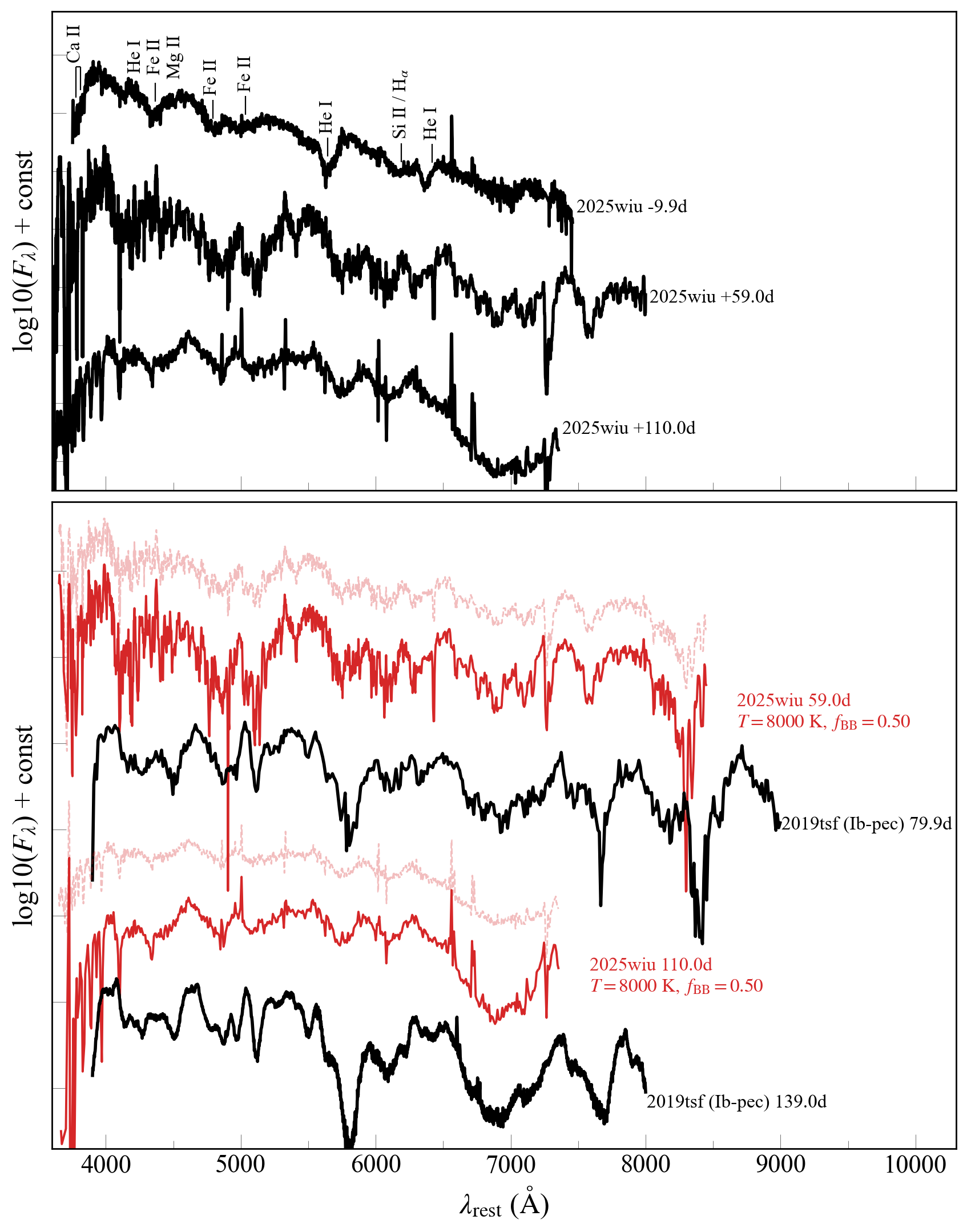}
        \end{overpic}
        \caption{Top: Comparison of the spectra of SN~2025wiu before, during, and after the post-maximum bump. Bottom: Comparison of the late-time spectra of SN~2025wiu before and after subtraction of an assumed blackbody continuum with those of SN~2019tsf. In the bottom panel, the original spectra of SN~2025wiu are shown in light red, while the blackbody-subtracted spectra are shown in red. A blackbody temperature of 8000 K is adopted, and the blackbody contribution $f_{\rm BB}$ is normalized to account for 50\% of the integrated spectral flux over the adopted wavelength range. The spectra are shifted vertically for clarity.}
        \label{fig:subtract_spec}
    \end{figure}
    
    The post-maximum spectra of SN~2025wiu and our selected SESN comparison sample are compared in Figure~\ref{fig:spec_compare_post_max}. {During the re-brightening bump phase of SN~2025wiu (left panel), its overall absorption features resemble those of most comparison SESNe, but are most similar to those of SN~2019tsf at this phase. These features show notably weaker equivalent widths than those of most comparison objects, suggesting the presence of an additional veiling continuum component. In particular, over the wavelength range from $\sim6400$ to $6700$~\AA, which covers $\mathrm{H}\alpha$ and \ion{He}{1}~$\lambda6678$, SN~2025wiu shows an apparent flux excess relative to most of the comparison spectra, with the sole exception of SN~2019tsf.} Interestingly, this flux excess vanishes entirely after the bump fades (right panel), implying a direct physical connection between the two features. Compared to other spectra at similar evolutionary phases, the spectrum of SN~2025wiu at $+110.0$ days shows remarkably weak absorption and emission features, reflecting its exceptionally slow spectral evolution and further suggesting continuous injection from an additional energy source. A weak emission-like feature near 6300~\AA\ is also visible in this spectrum and may be associated with the nebular [O~I] $\lambda\lambda6300,6364$ doublet.

    %A similra structure is also present in the spectra around maximum light (see Figure~\ref{fig:spec_compare_around_max}). This apparent excess may result from differences in the strength or velocity of nearby \ion{Si}{2} absorption, or from a complex blend of \ion{Si}{2} and \(\mathrm{H}\alpha\), which could reshape the local spectral profile and leave relatively more flux in this wavelength region.
 
   {To better illustrate the spectral evolution across the re-brightening bump, we directly compare the spectra in the upper panel of Figure~\ref{fig:subtract_spec}. Given the limited spectroscopic sampling, we used the $-9.9$ day spectrum to represent the pre-bump phase. We note that \citet{Lai_2026} presented a spectrum at $+24$ days after the primary maximum, which would be better suited to illustrating the spectral changes before and after the bump. The absorption features become stronger from the primary maximum to the bump, but show little further evolution after the bump. When combined with the $+24$ day spectrum presented by \citet{Lai_2026}, these spectra more fully illustrate the appearance and disappearance of the feature over $\sim6400$--$6700$\,\AA.} To illustrate the possible effect of an additional continuum component on the late-time spectra, we subtract an assumed blackbody spectrum from each observed spectrum in linear flux space. The resulting spectra are shown in the lower panel of Figure~\ref{fig:subtract_spec}. A blackbody temperature of 8000\,K is adopted, and the blackbody flux is normalized to approximately 50\% of the integrated spectral flux; this fraction is motivated by the contribution of the bump component to the total luminosity inferred from the light-curve analysis in Section~\ref{sec_6}. We emphasize that the blackbody assumption and the adopted temperature are illustrative rather than results of spectral fitting; nevertheless, after subtraction, the absorption and emission features have strengths broadly comparable to those observed in other SNe~Ib. At $+110$\,days, the absorption and emission features remain relatively weak, possibly because of residual contamination from the host galaxy that was not removed.

\section{Light-curve Analysis} \label{sec_5}

In this section, we investigate the ejecta properties and possible power source(s) of SN~2025wiu through light-curve modeling.

\subsection{Analytical Estimates of $^{56}$Ni} \label{Ni56}

One of the most widely used methods for estimating the synthesized $^{56}$Ni mass is the Arnett rule \citep{Arnett_1980,Arnett_1982}, which states that the instantaneous heating rate from the radioactive decay chain
$^{56}$Ni $\rightarrow$ $^{56}$Co $\rightarrow$ $^{56}$Fe equals the bolometric luminosity at maximum light:

\begin{equation}
\frac{L_{\rm peak}}{M_{\rm Ni}}
=
(\epsilon_{\rm Ni}-\epsilon_{\rm Co})
e^{-t_{\rm peak}/t_{\rm Ni}}
+
\epsilon_{\rm Co}
e^{-t_{\rm peak}/t_{\rm Co}},
\label{Eq:Arnett_Ni}
\end{equation}

where $\epsilon_{\rm Ni}=3.9\times10^{10}$ erg\,g$^{-1}$\,s$^{-1}$,
$\epsilon_{\rm Co}=6.8\times10^{9}$ erg\,g$^{-1}$\,s$^{-1}$,
$t_{\rm Ni}=8.8$\,days, and
$t_{\rm Co}=111.3$\,days are the specific heating rates and decay timescales of $^{56}$Ni and $^{56}$Co, respectively.

For SN~2025wiu, the first peak occurs at
$t_{\rm peak}\approx30$\,days after explosion with a bolometric luminosity of
$L_{\rm peak}=5.89\times10^{42}$ erg\,s$^{-1}$.
Assuming that the peak is powered solely by radioactive decay, Equation~(\ref{Eq:Arnett_Ni}) yields
$M_{\rm Ni}=0.47\,M_\odot$.
This value is substantially larger than those typically inferred for SNe~Ib, which are generally in the range of
$0.1$--$0.2\,M_\odot$ \citep{Lyman_2016,Anderson_2019}.

An independent estimate of $M_{\rm Ni}$ can be obtained by modeling the late-time radioactive tail.
At sufficiently late epochs, the ejecta become partially transparent to $\gamma$-rays, and the observed luminosity can be approximated as

\begin{equation}
L_{\rm obs}=f_{\rm dep}L_{\gamma},
\end{equation}
where $L_{\gamma}$ is the radioactive energy generation rate and $f_{\rm dep}$ is the deposition function, which describes the fraction of radioactive energy deposited in the ejecta.
In the optically thick limit, $f_{\rm dep}\approx1$, while in the optically thin limit $f_{\rm dep}\propto t^{-2}$.
A commonly adopted form is

\begin{equation}
f_{\rm dep}(t)
=
1-\exp\left[-\left(\frac{T_0}{t}\right)^2\right],
\end{equation}
where $T_0$ is the characteristic $\gamma$-ray escape timescale \citep{Sutherland_1984,Clocchiatti_1997,Haynie_2023}.

Fitting the linear decline phase of SN~2025wiu with $M_{\rm Ni}$ and $T_0$ as free parameters yields

\[
M_{\rm Ni}=0.13^{+0.05}_{-0.02}\,M_\odot
\]

and

\[
T_0=173^{+44}_{-41}~{\rm days}.
\]

The large discrepancy between the nickel masses inferred from the peak and tail luminosities is not unique to SN~2025wiu and has been reported for many SESNe \citep{Afsariardchi_2021, Rodriguez_2024}.
Part of this discrepancy may arise from limitations of the Arnett rule itself.
For example, \cite{Dessart_2015a} found that the Arnett rule can overestimate the true $^{56}$Ni mass by approximately 50\% in numerical simulations. However, even accounting for such a correction, the discrepancy in SN~2025wiu remains substantial.

\cite{Khatami_2019} relaxed the assumption of a self-similar internal energy profile adopted in the original Arnett model and derived the relation

\begin{equation}
L_{\rm peak}
=
\frac{2}{(\beta t_{\rm peak})^2}
\int_{0}^{\beta t_{\rm peak}}
\tau L_{\rm heat}(\tau)\,d\tau,
\end{equation}
where $\beta$ is a dimensionless parameter that can be calibrated using numerical simulations or observations \citep{fang2026}. Assuming that the tail-derived nickel mass provides a more reliable estimate of the radioactive content, we find that reproducing the first peak of SN~2025wiu requires
$\beta \approx 0.3$.
This is in stark contrast to the values of order unity obtained in the numerical calculations of \cite{Khatami_2019}.
The discrepancy suggests that the radioactive energy implied by the tail is insufficient to power the first light-curve peak, and it is difficult to be recovered by varying the concentration of the heating source or the composition in the ejecta. 

This result indicates that at least one of the following assumptions may break down for SN~2025wiu:

\begin{enumerate}
\item the first peak and late-time tail are both powered by radioactive decay of $^{56}$Ni; or
\item the ejecta opacity remains approximately constant, which is required by the underlying analytical framework.
% the ejecta opacity evolves significantly with time, which violates the approximately constant-opacity assumption underlying the analytical framework.
\end{enumerate}

Furthermore, the discrepancy may be even larger than estimated above.
The late-time luminosity could be contaminated by flux associated with the second bump in the light curve, causing the tail-derived nickel mass to be overestimated.
Consequently, the value
$M_{\rm Ni}=0.127\,M_\odot$
should be regarded as an upper limit.

\subsection{Semi-analytical Light-Curve Modeling}
\begin{figure}[htbp]
    \centering
    \begin{overpic}[width=0.485\textwidth]{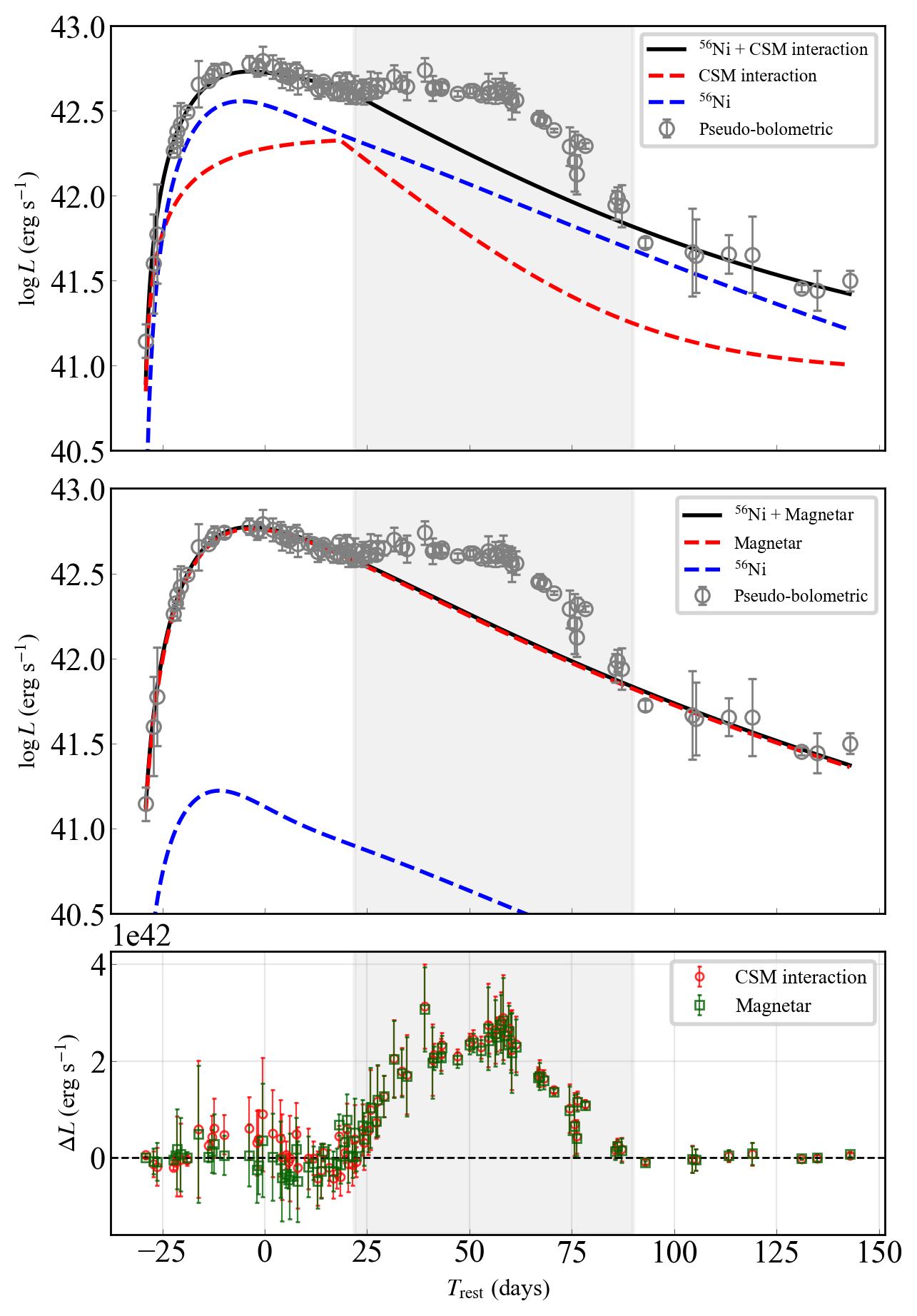}
    \end{overpic}
    \caption{Light-curve fits with the $^{56}$Ni+CSM interaction model (top) and the $^{56}$Ni+magnetar model (middle). Grey open circles show the pseudo-bolometric observations, and black solid lines show the total model luminosities. The contributions from radioactive decay and CSM interaction or magnetar power are shown as blue and red dashed lines, respectively. The fits are restricted to the first peak and the late-time decline, excluding the luminous post-maximum bump within the shaded region. The bottom panel shows the luminosity residuals for the CSM interaction model (red circles) and the magnetar model (green squares), illustrating the excess emission associated with the bump.}
    \label{fig:LC_fitting_combine}
\end{figure}

The analytical estimates presented above suggest that radioactive decay alone may be insufficient to account for the primary peak and late-time tail of SN~2025wiu under the adopted assumptions. To isolate the post-maximum bump, we fit these phases, excluding the bump, with radioactive decay combined with either CSM interaction or magnetar spin-down using \texttt{Redback} \citep{Sarin_2024}. The model descriptions, parameter settings, fitting procedure and results are presented in Appendix~\ref{sec:model}. Other mechanisms, such as fallback accretion (\citealt{Chan_2018,Moriya_2019}) and a combination of multiple processes, may produce similar luminosity evolution and cannot be ruled out with the current dataset.

{Both models reproduce the primary peak and late-time tail,
as shown in Figure~\ref{fig:LC_fitting_combine}.
The CSM-interaction model gives $M_{\rm Ni}\approx0.2\,M_\odot$
and $M_{\rm CSM}\approx0.1\,M_\odot$.
These values should be regarded as illustrative, model-dependent
estimates because the adopted model may not describe the actual
CSM geometry or optical depth.
For the magnetar model, the best-fitting solution is dominated
by magnetar power. The available data do not reliably constrain the relative contributions of radioactive decay and the additional energy source. Therefore, this does not imply a small synthesized nickel mass. The two models provide illustrative baselines, and the available light-curve data do not clearly distinguish between these mechanisms.}

{Subtracting the fitted baselines yields bump components with similar overall profiles, as shown in the lower panel of Figure~\ref{fig:LC_fitting_combine}. We discuss the properties and possible origins of the bump based on these fitting results below.}

\section{Bump Properties and Potential Energy Sources} \label{sec_6}
{Based on the bump component extracted using the $^{56}$Ni+CSM interaction ($^{56}$Ni+magnetar) baseline,} we derive the following bump properties. The time interval between the maximum and the bump peak is $\sim50\,{\rm days}$. The overall light-curve profile is relatively symmetric, and the peak luminosity and duration above half-maximum are $\sim2.44\,(2.37)\times10^{42}\,{\rm erg/s}$ and $\sim 41.9\,(42.2)\,{\rm days}$, respectively. The bump component contributes approximately half of the total bolometric luminosity. The total energy of $\sim9.4\,(9.2)\times10^{48}\,{\rm erg}$ is released. In the following, we discuss the viability of several alternative mechanisms as the underlying power source based on the derived properties of the bump.
    
The combination of shock cooling and radioactive decay is unlikely to explain the two peaks of SN~2025wiu, given the broad first peak and the $\sim50$\,day peak separation, both substantially longer than typical shock-cooling timescales in SESNe \citep[e.g.,][]{Zhao_2026}.

A double $^{56}$Ni distribution, proposed for SN~2005bf \citep{Tominaga_2005}, can produce two peaks powered by radioactive material in the outer and inner ejecta \citep{Orellana_2022a}. However, reproducing SN~2025wiu requires strong outward mixing of $^{56}$Ni, and its $\sim50$\,day peak separation approaches the model limit \citep[see Figure~2 in][]{Orellana_2022a}. This radioactive-only scenario also cannot simultaneously explain the primary peak and late-time tail (see subsection~\ref{Ni56}).

Other possible powering mechanisms include magnetar flares, fallback accretion, and hidden ejecta--CSM interaction. If the energy is deposited beneath the ejecta photosphere, its escape is delayed by radiative diffusion, characterized by an effective timescale $t_{\rm d}$ \citep[e.g.,][]{Arnett_1982}. Our semi-analytical fits with magnetar and CSM-interaction contributions give ejecta masses of $\sim3.4~M_\odot$ and $\sim6.3~M_\odot$, respectively. Adopting the corresponding fitted expansion velocities of $\sim8000$ and $\sim11000$ km s$^{-1}$ and an opacity of $\kappa=0.06$ cm$^2$ g$^{-1}$, we estimate effective diffusion timescales of approximately 23.3 and 27.0 days, respectively, using $t_{\rm d}=\sqrt{3\kappa M_{\rm ej}/(4\pi v_{\rm ej}c)}$.

For the scenario of a magnetar flare, since a typical flare timescale is exceptionally short ($\Delta t_{\mathrm{flare}} \ll t_{\mathrm{d}}$), the duration of the resulting light-curve feature is determined by the ejecta diffusion. Given our estimated $t_{\mathrm{d}}$ of $20$--$30$ days at this stage, a single flare may have difficulty reproducing the observed bump with the duration of $\sim 42$\,days. Multiple flares could prolong the emission, although their timing and energy would need to produce a smooth, single bump.

Standard fallback accretion declines as $\dot{M}\propto t^{-5/3}$ \citep{Chevalier_1989}, and diffusion over $\sim20$--$30$ days may not sufficiently smooth this asymmetric input to reproduce the observed symmetric bump. The bump peak at $\sim50$ days after maximum also suggests delayed energy injection, which is not readily explained by standard single-star models, although accretion from a binary companion could provide delayed input for suitable orbital configurations.

Hidden ejecta--CSM interaction could explain the bump without prominent narrow lines. In an asymmetric configuration, the interaction region may be buried beneath the ejecta photosphere. Alternatively, an optically thin, spherical CSM may produce narrow lines that are intrinsically weak or diluted by the SN continuum. An appropriate CSM geometry and density profile might reproduce the $\sim42$\,day duration, although further modeling is needed.

If the bump peak at $\sim80$ days after explosion mainly reflects the ejecta travel time, adopting a velocity of $8000$ km s$^{-1}$ gives a characteristic interaction radius of $\sim5\times10^{15}$ cm. For assumed CSM velocities of 100--1000 km s$^{-1}$, this corresponds to material ejected $\sim2$--$20$ years before explosion. The bump energy of $\sim10^{49}$ erg implies an interacting CSM mass of $\sim0.03~M_\odot$, assuming a relative velocity of $8000$ km s$^{-1}$ and a radiative efficiency of 0.5. These estimates are illustrative and do not constrain the geometry.

Further observational tests are therefore needed. Narrow or intermediate-width emission lines may emerge once the interaction region becomes exposed (e.g., \citealt{Pyykkinen_2026}), while radio observations can provide independent constraints on the surrounding material (e.g., \citealt{Zenati_2025}). No narrow emission lines are detected in our spectrum at $+110$ days. This could reflect continued obscuration or intrinsically weak line emission. We also inspected public archival data from the ASKAP RACS survey \citep{McConnell_2020}, which observed the field at approximately $+30$ and $+130$ days. No new radio source was detected at the SN position.

\section{Conclusion} \label{sec_7}

In this work, we present extensive photometric and spectroscopic observations of the peculiar SN Ib~2025wiu discovered by the WFST. SN~2025wiu is a luminous and slowly evolving event characterized by a prominent post-maximum bump in its light curves. Our photometric monitoring spans from $\sim1$ day after explosion to $\sim310$ days after maximum, providing detailed coverage of its evolution with a nearly daily cadence, including extensive $u$-band observations. The early light curve shows a possible shock-cooling excess consistent with an analytic model for a compact progenitor with a radius of approximately $4\,R_\odot$.

Several observational properties distinguish SN~2025wiu from typical SNe Ib. First, its luminous and slowly evolving main peak and late-time tail cannot be adequately explained by radioactive decay alone, indicating the presence of an additional energy source. Second, a prominent bump begins to emerge at $\sim23$ days after maximum and peaks at $\sim50$ days. The bump has a peak luminosity of $\sim2.4\times10^{42}\,\mathrm{erg\,s^{-1}}$ and a characteristic duration of $\sim42$ days, indicating additional energy input during the post-maximum evolution. Third, while the optical spectra of SN~2025wiu broadly resemble those of typical SNe Ib, their evolution is exceptionally slow, with persistently weak line features and no narrow emission lines detected even at late phases.

By modeling the main peak and late-time light curve with radioactive decay combined with either magnetar spin-down or CSM interaction, we obtain two alternative baselines that yield similar residual bump profiles. We then investigate the origin of this excess emission under several physical scenarios. A double-$^{56}$Ni distribution has difficulty reproducing the observed bump morphology. Magnetar flares and fallback accretion may in principle produce additional energy input, but explaining the observed light-curve structure requires additional assumptions, such as multiple episodic flares or accretion from a binary companion. Delayed CSM interaction remains a possible explanation for the bump, although the available data do not uniquely establish this mechanism.

The physical origin of such peculiar SNe Ib remains uncertain, in part because only a limited number of well-observed events are currently known. Ongoing and next-generation optical time-domain surveys, including WFST and the Vera C. Rubin Observatory's Legacy Survey of Space and Time (VRO-LSST), will substantially enlarge the sample and provide the systematic, high-cadence observations needed to investigate the diversity of their energy sources and progenitor environments.

%% Please use the acknowledgment and contribution environments. This will 
%% be anonomyized when the "anonymous" style option is used. 
\begin{acknowledgments}
This work was supported by the National Natural Science Foundation of China (Grant No. 12393811), the National Key R\&D Program of China (Grant No. 2023YFA1608100), and the Strategic Priority Research Program of the Chinese Academy of Science (Grant No. XDB0550300). Z.Y.L. acknowledges support from the Postdoctoral Fellowship Program of the China Postdoctoral Science Foundation (grant No. GZC20261722). J.J. acknowledges support from the Japan Society for the Promotion of Science (JSPS) KAKENHI grants JP22K14069. K.M. acknowledges support from the JSPS KAKENHI grant JP24KK0070 and 23H04894. K.M. and H.K. acknowledge support from the JSPS bilateral JPJSBP120229923. H.K. was funded by the Research Council of Finland projects 358691 and 358692. L.G. acknowledges financial support from MCIN and AEI 10.13039/501100011033 under project PID2023-151307NB-I00. B.W. is supported by the National Natural Science Foundation of China (No 12225304), the Yunnan Revitalization Talent Support Program (Yunling Scholar Project), the Yunnan Science and Technology Program (Nos 202501AS070005, 202605AS350010 and 202601BC070011).

The Wide Field Survey Telescope (WFST) is a joint facility of the University of Science and Technology of China, Purple Mountain Observatory. We appreciate the members of the WFST operation and maintenance team for their support. 

We acknowledge the support of the staff of the LJT. The LJT is supported by the Chinese Academy of Sciences (CAS) and the People's Government of Yunnan Province, and operated and managed by Yunnan Observatories, CAS.

Based on observations made with the Nordic Optical Telescope, owned in collaboration by the University of Turku and Aarhus University, and operated jointly by Aarhus University, the University of Turku and the University of Oslo, representing Denmark, Finland and Norway, the University of Iceland and Stockholm University at the Observatorio del Roque de los Muchachos, La Palma, Spain, of the Instituto de Astrofisica de Canarias. The NOT data were obtained under program ID P72-022.

The SALT data presented here were taken as part of Rutgers University program 2025-1-MLT-001 (PI: Jha).
\end{acknowledgments}

%\begin{contribution}
%%This section gives authors the space to recognize author contributions. The text inside this environment is NOT counted towards the total word quanta. At a minimum, manuscripts are expected to include this text:

%All authors contributed equally to the.

%% But authors are expected to provide more specific details, e.g. 
%%
%%SC was responsible for writing and submitting the manuscript.
%%WWM came up with the initial research concept and edited the manuscript.
%%OTS obtained the funding and edited the manuscript.
%%EBF provided the formal analysis and validation. He also edited the manuscript.
%%GEH Supervised the undergraduates, wrote the software and administers the project github and Zenodo repositories.
%%
%% Authors can use the Contributor Role Taxonomy (CRediT) at
%% for ideas on how write a good statement tailored to their needs.

%\end{contribution}

\facilities{WFST:2.5m, NOT, YAO:2.4m, ARC, SALT, GTC, Hale, PO:1.2m, ATLAS}

\software{
\texttt{Astropy} \citep{2013A&A...558A..33A,2018AJ....156..123A,2022ApJ...935..167A}, \texttt{Matplotlib} \citep{Hunter:2007}, SciPy \citep{2020SciPy-NMeth}, \texttt{NumPy} \citep{harris2020array}, \texttt{Astrometry.net} \citep{astrometry.net}, \texttt{SWarp} \citep{SWarp}, \texttt{Source Extractor} \citep{SExtractor}, \texttt{HOTPANTS} \citep{hotpants},  \texttt{Photutils} \citep{Photutils}, \texttt{dustmaps} \citep{Green2018}, \texttt{PypeIt} \citep{pypeit:joss_pub,pypeit:zenodo}, \texttt{pyKOSMOS} \citep{PyKOSMOS}, \texttt{Dynesty} \citep{Speagle_2020}, \texttt{Redback} \citep{Sarin_2024}, \texttt{Bilby} \citep{Ashton_2019}
}


\appendix
\section{Light-curve Modeling with Additional Energy Sources} \label{sec:model}
Here we describe the parameter settings and fitting procedure for the two baseline models presented in the main text. Both models are fitted to the primary peak and late-time decline. We use the bolometric models implemented in
\texttt{Redback} \citep{Sarin_2024}. The radioactive component is described by
\texttt{arnett\_\allowbreak bolometric}, with the nickel mass parameterized as $M_{\rm Ni}=f_{\rm Ni}M_{\rm ej}$. The nickel fraction is allowed to vary in both models.

All times are evaluated in the rest frame. The model time since explosion is calculated using the estimated explosion epoch and an additional parameter, $t_{\rm shift}$, which accounts for uncertainty in the explosion epoch.

For both models, we fix the optical opacity to $\kappa=0.06$ cm$^2$ g$^{-1}$ \citep{Maeda_2003,Valenti_2011} to reduce its degeneracy with the ejecta mass. For the radioactive component, we adopt $\kappa_{\gamma,\rm Ni}=0.025$ cm$^2$ g$^{-1}$ \citep{Wygoda_2019}. We use a Gaussian prior on the ejecta velocity, centered at $8000$ km s$^{-1}$ with a standard deviation of $1000$ km s$^{-1}$, and reject non-positive velocities. This prior is guided by our uncertain Fe~II $\lambda5169$ measurement near maximum and the velocity of $7690\pm190$ km s$^{-1}$ measured at $+24$ days by \citet{Lai_2026}.

\subsection{CSM Interaction plus Radioactive Decay}

{For the interaction component, we use \texttt{csm\_\allowbreak interaction\_\allowbreak bolometric} and add its emergent luminosity to that of \texttt{arnett\_\allowbreak bolometric}. The interaction model follows the one-dimensional, spherically symmetric framework of \citet{Chatzopoulos_2013}, including heating from forward and reverse shocks and radiative diffusion through optically thick CSM. The two components share the ejecta mass, ejecta velocity, and optical opacity. The CSM is characterized by its mass, $M_{\rm CSM}$, inner and outer radii, $R_{\rm inner}$ and $R_{\rm outer}$, and density-profile index, $s$. We fix the inner and outer ejecta density indices to $\delta=1$ and $n=12$, respectively, and the radiative conversion efficiency to $\epsilon=0.5$. The adopted priors are summarized in Table~\ref{tab:csm_fitting}. The inferred CSM parameters are effective, model-dependent quantities: the assumptions may not describe either a buried, asymmetric interaction region \citep[e.g.,][]{Smith_2016a} or optically thin CSM. A detailed treatment would require models appropriate to the optical depth and, for asymmetric configurations, multidimensional calculations \citep[e.g.,][]{Suzuki_2019}, beyond the scope of this work.}

\subsection{Magnetar plus Radioactive Decay}
For the magnetar component, we use \texttt{basic\_\allowbreak magnetar\_\allowbreak powered\_\allowbreak bolometric}, parameterized by the total spin-down energy, $E_{\rm total}$, and characteristic timescale, $t_{\rm p}$. These are converted to the spin period and magnetic field required by the model, without additional priors on the derived quantities. The magnetar and radioactive components share $M_{\rm ej}$, $v_{\rm ej}$, and $\kappa$. Their luminosities are calculated separately, including diffusion and gamma-ray leakage, and then added. The effective magnetar gamma-ray opacity, $\kappa_{\gamma,\rm mag}$, is allowed to vary. The adopted priors and fixed values are listed in Table~\ref{tab:priors}. The relative contributions of the two components cannot be reliably constrained; the magnetar model serves only as an illustrative baseline and does not establish a magnetar origin.

% Required in the preamble:
% \usepackage{graphicx}

\begin{table*}[!t]
\noindent
% ---------------- Left table ----------------
\begin{minipage}[t]{0.49\textwidth}
\vspace{0pt}

\caption{Parameters, priors, and posterior constraints of the
CSM interaction+$^{56}$Ni model.}
\label{tab:csm_fitting}

\begingroup
\normalsize
\setlength{\tabcolsep}{2.5pt}
\setlength{\parindent}{0pt}
\par\noindent
\makebox[\linewidth][l]{%
\resizebox{\linewidth}{!}{%
\begin{tabular}{@{}lcccccc@{}}
\hline
Parameter & Distribution & Min & Max & Mean & Std & Posterior \\
\hline
$M_{\rm ej}/M_\odot$
& Log-uniform & 0.1 & 10 & -- & -- & $6.3_{-0.6}^{+0.6}$ \\
$v_{\rm ej}/(10^3~{\rm km~s^{-1}})$
& Gaussian & -- & -- & 8 & 1 & $10.9_{-0.6}^{+0.6}$ \\
$M_{\rm CSM}/M_\odot$
& Log-uniform & 0.001 & 1 & -- & -- & $0.10_{-0.08}^{+0.4}$ \\
$R_{\rm inner}/(10^{15}~{\rm cm})$
& Log-uniform & 1 & 10 & -- & -- & $8.8_{-0.9}^{+0.7}$ \\
$R_{\rm outer}/(10^{16}~{\rm cm})$
& Log-uniform & 1 & 100 & -- & -- & $7_{-5}^{+10}$ \\
$s$
& Uniform & 1.2 & 1.8 & -- & -- & $1.67_{-0.03}^{+0.03}$ \\
$f_{\rm Ni}$
& Log-uniform & 0.001 & 1 & -- & -- & $0.034_{-0.002}^{+0.003}$ \\
$t_{\rm shift}/{\rm day}$
& Uniform & $-5$ & 0.7 & -- & -- & $0.26_{-0.08}^{+0.07}$ \\
$\kappa/({\rm cm^2~g^{-1}})$
& Fixed & -- & -- & 0.06 & -- & -- \\
$\kappa_{\gamma,\rm Ni}/({\rm cm^2~g^{-1}})$
& Fixed & -- & -- & 0.025 & -- & -- \\
$\epsilon$
& Fixed & -- & -- & 0.5 & -- & -- \\
$n$
& Fixed & -- & -- & 12 & -- & -- \\
$\delta$
& Fixed & -- & -- & 1 & -- & -- \\
$M_{\rm Ni}/M_\odot$
& Derived & -- & -- & -- & -- & $0.21_{-0.02}^{+0.02}$ \\
\hline
\end{tabular}%
}%
}
\par
\endgroup

\smallskip
{\footnotesize\raggedright
\setlength{\parindent}{0pt}
\textit{Note.}---Posterior values are medians with
16th--84th percentile credible intervals. The nickel mass is derived as
$M_{\rm Ni}=f_{\rm Ni}M_{\rm ej}$.\par}
\end{minipage}%
\hfill
% ---------------- Right table ----------------
\begin{minipage}[t]{0.49\textwidth}
\vspace{0pt}

\caption{Parameters, priors, and posterior constraints of the
magnetar+$^{56}$Ni model.}
\label{tab:priors}

\begingroup
\normalsize
\setlength{\tabcolsep}{2.5pt}
\setlength{\parindent}{0pt}
\par\noindent
\makebox[\linewidth][l]{%
\resizebox{\linewidth}{!}{%
\begin{tabular}{@{}lcccccc@{}}
\hline
Parameter & Distribution & Min & Max & Mean & Std & Posterior \\
\hline
$M_{\rm ej}/M_\odot$
& Log-uniform & 0.1 & 10 & -- & -- & $3.4_{-0.5}^{+0.6}$ \\
$v_{\rm ej}/(10^3~{\rm km~s^{-1}})$
& Gaussian & -- & -- & 8 & 1 & $8.1_{-0.9}^{+0.9}$ \\
$E_{\rm total}/(10^{48}~{\rm erg})$
& Log-uniform & 1 & $10^{3}$ & -- & -- & $80_{-10}^{+10}$ \\
$t_{\rm p}/{\rm day}$
& Log-uniform & 0.1 & $10^{3}$ & -- & -- & $100_{-20}^{+30}$ \\
$f_{\rm Ni}$
& Log-uniform & 0.001 & 1 & -- & -- & $0.004_{-0.002}^{+0.006}$ \\
$\kappa_{\gamma,\rm mag}/({\rm cm^2~g^{-1}})$
& Log-uniform & $10^{-4}$ & 1 & -- & -- & $0.018_{-0.003}^{+0.003}$ \\
$t_{\rm shift}/{\rm day}$
& Uniform & $-5$ & 5 & -- & -- & $-1.3_{-0.2}^{+0.2}$ \\
$\kappa/({\rm cm^2~g^{-1}})$
& Fixed & -- & -- & 0.06 & -- & -- \\
$\kappa_{\gamma,\rm Ni}/({\rm cm^2~g^{-1}})$
& Fixed & -- & -- & 0.025 & -- & -- \\
$M_{\rm Ni}/M_\odot$
& Derived & -- & -- & -- & -- & $0.012_{-0.007}^{+0.02}$ \\
\hline
\end{tabular}%
}%
}
\par
\endgroup

\smallskip
{\footnotesize\raggedright
\setlength{\parindent}{0pt}
\textit{Note.}---Posterior values are medians with
16th--84th percentile credible intervals. The nickel mass is derived as
$M_{\rm Ni}=f_{\rm Ni}M_{\rm ej}$.\par}
\end{minipage}
\end{table*}

\subsection{Fitting Procedure}

We adopt a Gaussian likelihood in logarithmic luminosity, using the reported uncertainties without additional weighting or an intrinsic-scatter term. The posterior distributions are sampled with \texttt{Dynesty} through \texttt{Bilby}, using 1200 live points and a stopping threshold of $\Delta\ln Z=0.1$. The CSM and magnetar models use internal time grids of 6400 and 3200 points, respectively; numerical sensitivity is checked by doubling the grid resolution at the maximum-likelihood sample. The corresponding posterior constraints of the CSM and magnetar models are summarized in Table~\ref{tab:csm_fitting} and Table~\ref{tab:priors}, respectively.

\section{Dataset}
In this Appendix, we present the supporting data and logs for our observational campaigns of SN~2025wiu. Table~\ref{tab:pho_summary} summarizes the 12 selected epochs of photometric follow-up observations, detailing the instruments and filter bands used. The associated uncorrected apparent magnitudes from our comprehensive monitoring are partially listed in Table~\ref{tab:pho_data_list}, while the complete, machine-readable dataset is hosted in the online version of the journal. Additionally, Table~\ref{tab:spec_summary} provides the detailed log for the seven epochs of spectroscopic observations, spanning from $\sim -10$\,days to $\sim 110$\,days relative to the $r$-band maximum, including information on the telescopes, grisms, and wavelength coverage.
\begin{table}[htbp]
\centering
\footnotesize
\begin{threeparttable}
    \renewcommand{\arraystretch}{1.5}
    \setlength{\tabcolsep}{10pt}
    \caption{\noindent\textbf{Summary of 12 epochs of photometric follow-up observations for SN~2025wiu.}} 
        \begin{tabular}{*{4}{c}}
            \toprule 
            MJD & Phase\tnote{1} (day) & Band  & Telescope/Instrument \\
            \midrule
            60985.01 &  36.98 & $HJK_s$ & NOT/NOTCam \\
            60989.10 &  40.88 & $griz$  & NOT/ALFOSC \\
            61009.00 &  59.89 & $HJK_s$ & NOT/NOTCam \\
            61026.02 &  76.14 & $griz$  & NOT/ALFOSC \\
            61035.98 &  85.65 & $griz$  & NOT/ALFOSC \\
            61064.85 & 113.24 & $griz$  & NOT/ALFOSC \\
            61070.91 & 119.02 & $griz$  & NOT/ALFOSC \\
            61074.83 & 122.77 & $HJK_s$ & NOT/NOTCam \\
            61083.51 & 131.05 & $gri$   & LJT/YFOSC \\
            61087.53 & 134.89 & $gri$   & LJT/YFOSC \\
            61095.52 & 142.52 & $g$     & LJT/YFOSC \\
            61095.87 & 142.87 & $griz$  & NOT/ALFOSC \\
            \bottomrule
        \end{tabular}
    \label{tab:pho_summary}
    \begin{tablenotes}
        \footnotesize
        \item[1] Days relative to the epoch of $r$-band maximum in the rest frame.
    \end{tablenotes}
\end{threeparttable}
\end{table}

\begin{table}
\centering
\begin{threeparttable}
    \renewcommand{\arraystretch}{1.1}
    \setlength{\tabcolsep}{10pt}
    \caption{\noindent\textbf{Partial photometric data for SN~2025wiu without correction for Milky Way extinction. The full version of this table is available in machine-readable format in the online journal.}}
    \label{tab:pho_data_list}

    \begin{tabular}{*{5}{c}}
        \toprule
        MJD &
        Phase\tnote{1} (day) &
        Band &
        Apparent Magnitude\tnote{2} &
        Telescope/Instrument \\
        \midrule

        \multicolumn{5}{c}{WFST observations} \\
        \midrule
        60915.89 & $-29.04$ & $g$   & $22.7 \pm 0.1$   & WFST \\
        60916.89 & $-28.09$ & $r$   & $21.85 \pm 0.07$ & WFST \\
        % 60917.88 & $-27.14$ & $g$   & $21.85 \pm 0.05$ & WFST \\
        % 60918.81 & $-26.26$ & $g$   & $21.6 \pm 0.1$   & WFST \\
        % 60922.83 & $-22.42$ & $g$   & $20.06 \pm 0.03$ & WFST \\
        \multicolumn{5}{c}{$\cdots$} \\

        \midrule
        \multicolumn{5}{c}{Follow-up observations} \\
        \midrule
        60985.01 & 36.98 & $J$     & $18.68 \pm 0.07$ & NOT/NOTCam \\
        60985.03 & 36.99 & $H$     & $19.1 \pm 0.1$   & NOT/NOTCam \\
        % 60985.05 & 37.01 & $K_s$   & $19.4 \pm 0.1$   & NOT/NOTCam \\
        % 60989.10 & 40.88 & $g$     & $19.76 \pm 0.08$ & NOT/ALFOSC \\
        % 60989.10 & 40.88 & $r$     & $19.00 \pm 0.04$ & NOT/ALFOSC \\
        \multicolumn{5}{c}{$\cdots$} \\

        \bottomrule
    \end{tabular}

    \begin{tablenotes}
        \footnotesize
        \item[1] Days relative to the epoch of $r$-band maximum in the rest frame.
        \item[2] All magnitudes are reported in the AB system.
    \end{tablenotes}
\end{threeparttable}
\end{table}

\begin{table}[htbp]
\centering
\footnotesize
\begin{threeparttable}
    \renewcommand{\arraystretch}{1.5}
    \setlength{\tabcolsep}{10pt}
    \caption{\noindent\textbf{Summary of seven epochs of spectroscopic observations for SN~2025wiu.}} 
        \begin{tabular}{*{5}{c}}
            \toprule 
            MJD & Phase\tnote{1} (day) & Range ($\text{\AA}$) & Telescope/Instrument & Disperser \\
            \midrule
            60935.9698 & -9.9 & 3700--7800 & SALT/RSS & PG0900 \\
            60941.3455 & -4.7 & 3500--9500 & P60/SEDM\tnote{2} & Triple Prism \\
            60953.3036 & 6.7 & 5500--10400 & P200/NGPS & Red Arm Grating\\
            60999.6302 & 50.9 & 3400--9100 & LJT/YFOSC & Grism \#3 \\
            61008.0792 & 59.0 & 3800--9400 & ARC 3.5m/KOSMOS & Blue \& Red Grisms\\
            61019.5565 & 70.0 & 3400--9100 & LJT/YFOSC & Grism \#3 \\
            61061.5000 & 110.0 & 3600--7700 & GTC/OSIRIS & R1000B \\
            \bottomrule
        \end{tabular}
    \begin{tablenotes}
        \footnotesize
        \item[1] Days relative to the epoch of $r$-band maximum in the rest frame.
        \item[2] This spectrum is from TNS \citep{P60_TNS}.
    \end{tablenotes}
    \label{tab:spec_summary}
\end{threeparttable}
\end{table}

%% For this sample we use BibTeX plus aasjournalv7.bst to generate the

\bibliography{main}{}
\bibliographystyle{aasjournalv7}



\end{document}